\documentclass[aps,prl,twocolumn,showpacs,amsmath,amssymb,superscriptaddress]{revtex4-1}
\usepackage{graphicx}
\usepackage{latexsym}
\usepackage{amsmath}
\usepackage{color}
\usepackage{dcolumn}
\usepackage{bm}
\usepackage{mathrsfs}
\usepackage{feynmf}
\usepackage{comment}
\usepackage{float}
\usepackage{tikz}
\usepackage[colorlinks,citecolor=blue,urlcolor=blue]{hyperref}

\newcommand{\beq}{\begin{equation}}
\newcommand{\eeq}{\end{equation}}
\newcommand{\ba}{\begin{array}{ccc}}
\newcommand{\ea}{\end{array}}
\newcommand{\nn}{\nonumber \\}
\def\bea{\begin{eqnarray}}
\def\eea{\end{eqnarray}}

\begin{document}

\title{A quantum dimer model for the pseudogap metal}

\author{Matthias Punk}
 \affiliation{Institute for Theoretical Physics, University of Innsbruck, 6020 Innsbruck, Austria}
 \affiliation{Institute for Quantum Optics and Quantum Information, Austrian Academy of Sciences, 6020 Innsbruck, Austria}
 \affiliation{Physics Department, Ludwig-Maximilians-Universit\"at M\"unchen, 80333 Munich, Germany}

\author{Andrea Allais}
\affiliation{Department of Physics, Harvard University, Cambridge MA 02138, USA}

\author{Subir Sachdev}
\affiliation{Department of Physics, Harvard University, Cambridge MA 02138, USA}
\affiliation{Perimeter Institute for Theoretical Physics, Waterloo, Ontario N2L 2Y5, Canada}

\date{\today}

\begin{abstract}
We propose a quantum dimer model for the metallic state of the hole-doped cuprates at low hole density, $p$.
The Hilbert space is spanned by spinless, neutral, bosonic dimers and spin $S=1/2$, charge $+e$ fermionic
dimers. The model realizes a `fractionalized Fermi liquid' with no symmetry-breaking and 
small hole pocket Fermi surfaces enclosing a total area
determined by $p$. Exact diagonalization, on lattices of sizes up to $8 \times 8$, shows 
anisotropic quasiparticle residue around the pocket Fermi surfaces. We discuss the relationship
to experiments.
\end{abstract}


\maketitle

\section{Significance}

{\bf
The most interesting states of the copper oxide compounds are not the superconductors with high critical temperatures. Instead, the novelty lies primarily in the higher temperature metallic ``normal'' states from which the superconductors descend. Here we develop a simple, intuitive model for
the physics of the metal at low carrier density, in the ``pseudogap'' regime. This model describes a novel metal which is similar in many respects
to simple metals like silver. However the simple metallic character co-exists with `topological order' and long-range quantum entanglement previously
observed only in exotic insulators or fractional quantum Hall states in very high magnetic fields. Our model is compatible with many recent observations,
and we discuss more definitive experimental tests.}

\section{Introduction}

\begin{figure}
\begin{center}
\includegraphics[width=0.8\linewidth]{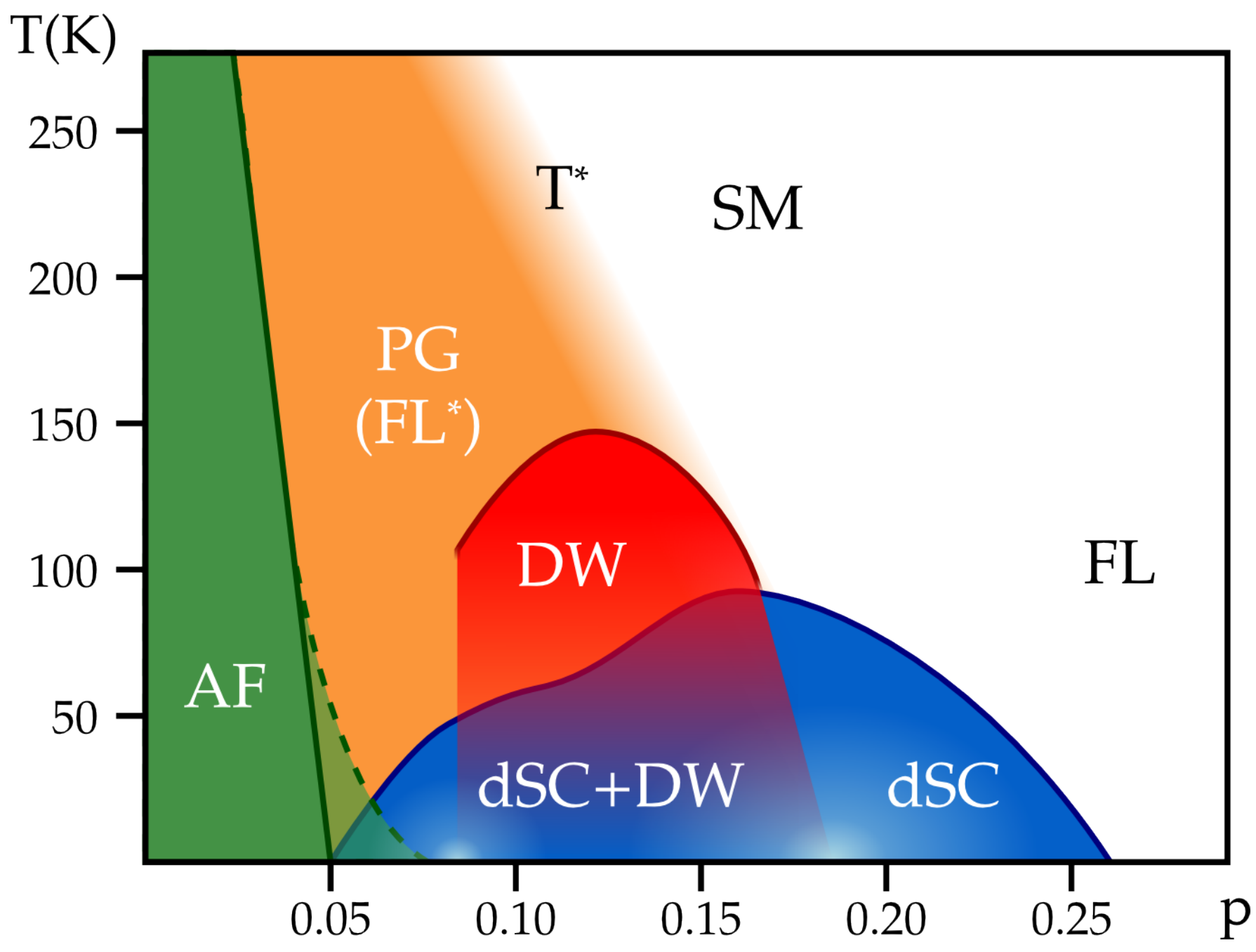}
\end{center}
\caption{Schematic phase diagram of hole-doped cuprates (apart from those with La doping) as a function
of temperature ($T$) and hole density ($p$). The antiferromagnetic (AF) insulator is present near $p=0$, and
the $d$-wave superconductor (dSC) is present below a critical temperature $T_c$. The pseudogap (PG) is present
for $T< T^\ast$ and acquires density-wave (DW) order at low $T$. The metallic states are the PG metal, the 
conventional Fermi liquid (FL) and the strange metal (SM). The dimer model of the present paper describes only
the PG metal as a fractionalized Fermi liquid (FL*).}
\label{fig:phasediag}
\end{figure}

The recent experimental progress in determining the phase diagram of the hole-doped Cu-based high temperature superconductors has highlighted the unusual and remarkable properties of the pseudogap (PG) metal --- see Fig.~\ref{fig:phasediag}. A characterizing feature of this phase is a depletion of the electronic density of states at the Fermi energy \cite{SB88,HA88}, anisotropically distributed in momentum space, that persists up to room temperature.

Attempts have been made to explain the pseudogap metal 
using thermally fluctuating order parameters; we argue below that such approaches are difficult to reconcile with recent 
transport experiments. Instead, we introduce a new microscopic model that realizes an alternative perspective \cite{DCSS15}, in which the pseudogap metal is a 
finite temperature ($T$) realization of a novel 
quantum state: the fractionalized Fermi liquid (FL*). We show that our model is consistent with 
key features of the pseudogap metal observed by both transport and spectroscopic probes.

The crucial observation that motivates our work is the tension between photoemission and transport experiments. In the cuprates, the hole density $p$ is conventionally measured relative to that of the insulating antiferromagnet (AF), which has one electron per site in the Cu $d$ band. Therefore, the hole density relative to a filled Cu band, with two electrons per site, is actually $1+p$. In fact, photoemission experiments at large hole-doping observe a Fermi surface enclosing an area determined by the hole density $1+p$ \cite{Plate95}, in agreement with the Luttinger relation. In contrast, in the pseudogap metal  a mysterious `Fermi arc' spectrum is observed \cite{ZXSRMP03,Shen05,PJ11}, with no clear evidence of closed Fermi surfaces. However, despite this unusual spectroscopic feature, transport measurements report vanilla Fermi-liquid properties, but associated with carrier density $p$, rather than $1+p$. The carrier density of $p$ was indicated directly in Hall measurements \cite{Ando04}, while other early experiments 
indicated suppression of the Drude weight \cite{Orenstein1990,Uchida91,LNW06}. While the latter could be compatible with a carrier density
of $1+p$ but with a suppressed kinetic term, the Hall measurements indicate the suppression of the Drude weight is more likely due to 
a small carrier density. Two recent experiments displaying Fermi liquid behavior at low $p$ are especially notable:
\begin{itemize}
\item The quasiparticle lifetime $\tau(\omega, T)$ determined from optical conductivity experiments \cite{Marel13} has the Fermi-liquid-like dependence $1/\tau \propto (\hbar \omega)^2 + (c \pi k_B T)^2$, with $c$ an order unity constant.
\item The in-plane magnetoresistance of the pseudogap \cite{MG14} is proportional to $ \tau^{-1} \left( 1 + b H^2 \tau^2 + \ldots\right)$ in an applied field $H$, where $\tau \sim T^{-2}$ and $b$ is a $T$-independent constant; this is Kohler's rule for a Fermi liquid. 
\end{itemize}

It is difficult to account for the nearly perfect Fermi-liquid-like $T$ dependence in transport properties of the pseudogap
in a theory in which a large Fermi surface of size $1+p$ \cite{AADCSS14} is disrupted by a thermally fluctuating order. In such a theory, we expect that transport should instead reflect the $T$ dependence of the correlation length of the order.

Moreover, a reasonable candidate for the fluctuating order has not yet been identified. The density wave (DW) order present at lower temperature in the pseudogap regime has been identified to have a $d$-form factor \cite{SSRLP13, Fujita14, Comin14, Forgan15, DFFDW15}, and its temperature dependence \cite{MHJ11, MHJ13, Ghiringhelli12, DGH12, Chang12, DLB13, Julien2015} indicates that it is unlikely to be the origin of the pseudogap present at higher temperature. Similar considerations apply to other fluctuating order models \cite{SKRMP03} based on AF or dSC. 

We are therefore led to
an alternative perspective \cite{DCSS15}, in which the pseudogap metal represents a new quantum state which could be stable down
to very low $T$, at least for model Hamiltonians not too different from realistic cuprate models. The observed low $T$ DW order is then
presumed to be an instability of the pseudogap metal \cite{MV12,comin13,neto13,DCSS14b,Mei14}. An early discussion \cite{Kotliar88} of the pseudogap metal proposed a state which was a doped spin liquid
with `spinon' and `holon' excitations fractionalizing the spin and charge of an electron: the spinon carries spin $S=1/2$ and is charge neutral,
while the holon is spinless and carries charge $+e$. However this state is incompatible with the sharp `Fermi arc' 
photoemission spectrum \cite{PJ11} around the diagonals of the Brillouin zone: the spin liquid has no sharp excitations with the quantum number of an electron, and so will only produce broad multi-particle continua in photoemission. 

Instead, we need a quantum state which has long-lived electron-like quasiparticles around a Fermi surface of size $p$, even though
such a Fermi surface would violate the Luttinger relation of a Fermi liquid. The fractionalized Fermi liquid (FL*) \cite{TSSSMV03} fulfills these requirements.

\section{Fractionalized Fermi Liquids}

The key to understanding the FL* state is the topological nature of the Luttinger relation for the area enclosed by the Fermi surface.
For the case of a conventional FL state, Oshikawa \cite{MO00} provided a non-perturbative proof of the Luttinger relation
by placing the system on a torus, and computing the response to a single flux quantum threaded through one of the holes of the torus.
His primary assumption about the many-body state was that its {\em only\/} 
low energy excitations were fermionic quasiparticles around a Fermi surface. This assumption then points to a route to obtaining
a Fermi surface of a different size \cite{TSMVSS04}: we need a metal which, in addition to the quasiparticle excitations around the
Fermi surface, has global topological excitations nearly degenerate with the ground state, similar to those found in insulating 
spin liquids \cite{NRSS91,XGW91}. In the context of the doped spin liquids noted earlier, we obtain a FL* state
when the holon and spinon bind to form a fermionic state
with spin $S=1/2$ and charge $+e$ (a possible origin of the binding is the attraction arising from the nearest-neighbor hopping), and there is a Fermi surface with quasiparticle excitations of this bound state \cite{Wen96,RK07,YQSS10} (other possibilities for the fate of this bound
have also been discussed \cite{Baskaran07}). 
Such a Fermi surface has long-lived electron-like quasiparticles and encloses an area determined by density $p$, and not $1+p$ \cite{RK08,YQSS10,LVVFL12,MPSS12}, just as required by observations in the pseudogap metal. Alternatively, a FL* phase can also
be obtained from Kondo lattice models \cite{NAPC89,BGG02}, but we shall not use this here.

Earlier studies have examined a number of phenomenological and path integral models of FL* theories of the pseudogap
\cite{RK07,RK08,YQSS10,LVVFL12,MPSS12} (and in an ansatz for the pseudogap \cite{YRZ06}). 
These models contain emergent gauge field excitations, which are need to provide the global topological states required to violate
the Luttinger relation of the FL state. But they also
include spurious auxiliary particle states which are only 
approximately projected out. The gauge field can undergo a crossover to confinement, but the present models
do not keep close track of lattice scale Berry phases which control the appearance of density wave order in the 
confining state \cite{NRSS89}. Here, we propose to overcome these difficulties by a new quantum dimer model which can realize a metallic state which is a FL*. 
This should open up studies of the photoemission spectrum, density wave instabilities, and crossovers to confinement at low $T$ in the
pseudogap metal.

\section{Quantum dimer models} 

Quantum dimer models \cite{DRSK88, EFSK90, RS89} have been powerful tools in uncovering the physics of spin liquid phases, and of their instabilities to conventional confining phases \cite{NRSS90, MVSS99, MS01}. Dimer models of doped spin liquids have also been studied \cite{DRSK88, DP08, Poilblanc12}, but all of these involve doping the insulating models by {\em monomers\/} which carry charge $+e$, but no spin. Here we introduce a new route to doping, in which the dopants are {\em dimers\/}, carrying both charge {\em and} spin.

\begin{figure}
\begin{center}
\includegraphics[width=2.5in]{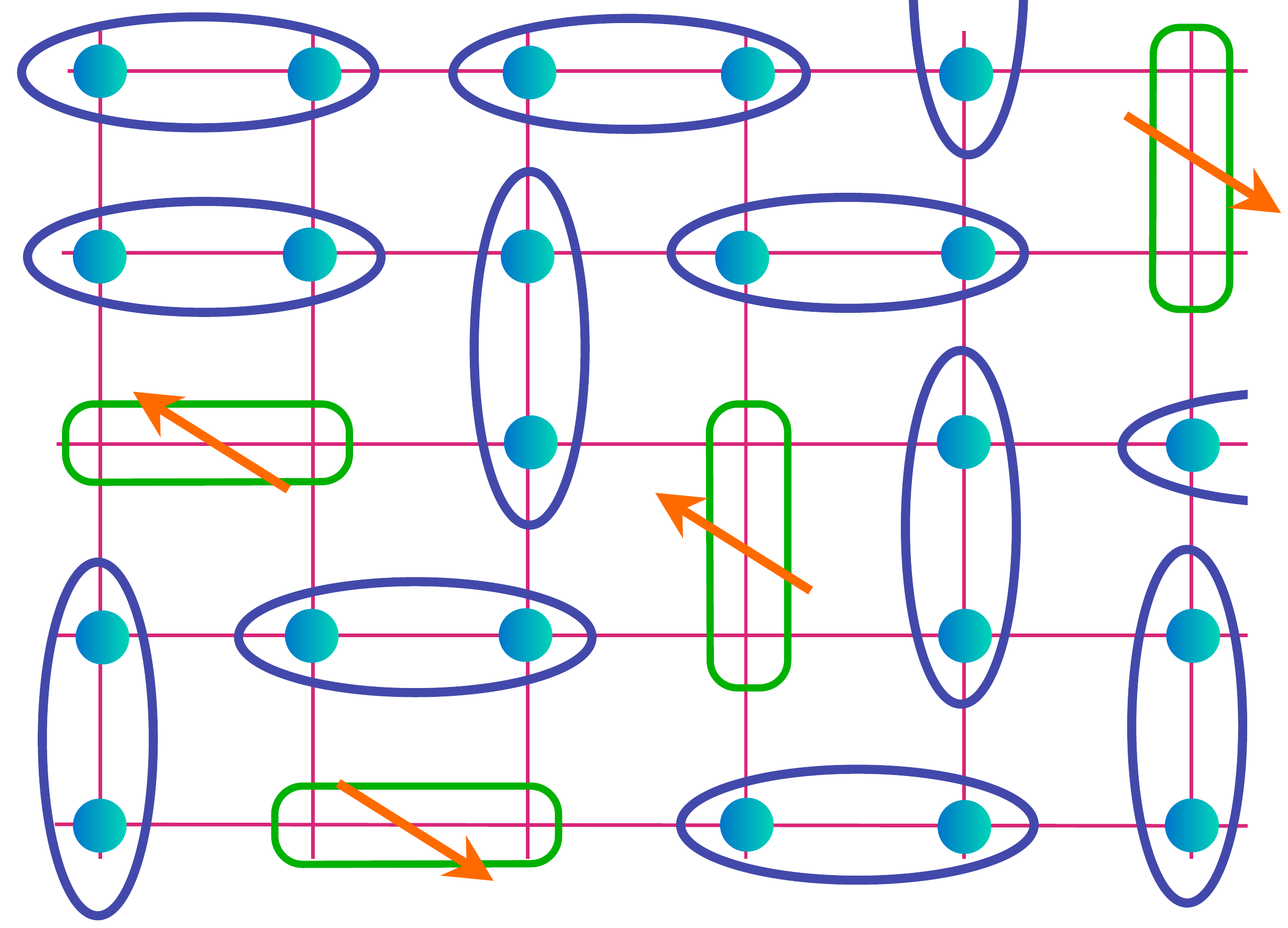}
\end{center}
\caption{(Color online) A typical dimer configuration identifying a state in the Hilbert space. The blue ellipses are the bosons $D_{i\eta}$ which are spinless and neutral.
The green rectangles are the fermions $F_{i\eta\alpha}$ which carry spin $S=1/2$ and charge $+e$. The density of the $F_{i\eta\alpha}$ dimers is $p$.}
\label{fig:hilbert}
\end{figure}

The Hilbert space of our dimer model is spanned by the close-packing coverings of the square lattice with two species of dimers (see Fig.~\ref{fig:hilbert}), with an additional two-fold spin degeneracy of the second species. It can be mapped by an appropriate similarity transform \cite{DRSK88} to a truncation of the Hilbert space of the $t$-$J$ model.

The first species of dimers are bosons, $D_{i \eta}$, which reside on the link connecting the square lattice site $i \equiv (i_x, i_y)$ to the site $i + \hat{\eta}$, where $\hat{\eta} = \hat{x} \equiv (1,0)$ or $\hat{y} \equiv (0,1)$. These are the same as the dimers in the Rokhsar-Kivelson (RK) model \cite{DRSK88}, to which our model reduces at zero doping. When connecting to the Hilbert space of the $t$-$J$ model, each boson maps to a pair of electrons in a spin-singlet state:
\beq
D_{i \eta}^\dagger \left| 0 \right\rangle ~\Rightarrow~ \Upsilon_{i \eta} (c_{i \uparrow}^\dagger c_{i +\hat{\eta},\downarrow}^\dagger + c_{i +\hat{\eta} ,\uparrow}^\dagger c_{i \downarrow}^\dagger)\left| 0 \right\rangle/\sqrt{2}  , \label{D1}
\eeq
where $c_{i \alpha}$ is the electron annihilation operator on Cu site $i$ with spin $\alpha = \uparrow, \downarrow$, and $\left| 0 \right\rangle$ is the empty state with no dimers or electrons. The phase factors $\Upsilon_{i \eta}$ depend upon a gauge choice: for the choice made by RK,
$\Upsilon_{i y}=1$ and $\Upsilon_{i x} = (-1)^{i_y}$.

The second species of dimers are {\it fermions\/}, $F_{i \eta \alpha}$ with $\alpha=\uparrow,\downarrow$, which carry spin $S=1/2$ and charge $+e$ relative to the half-filled insulator, and are present with a density $p$. Each fermionic dimer maps to a bound state of a holon and a spinon, which we take to reside on a bonding orbital between nearest-neighbor sites:
\beq
F_{i \eta \alpha}^\dagger \left| 0 \right\rangle ~\Rightarrow~ \Upsilon_{i \eta} (c_{i \alpha}^\dagger  + c_{i +\hat{\eta} ,\alpha}^\dagger )\left| 0 \right\rangle/\sqrt{2}. \label{F1}
\eeq

In a three-band model \cite{Emery87,Berciu}, the state $F_{i \eta \alpha}^\dagger \left| 0 \right\rangle$  can be identified with the $S=1/2$ state of a hole delocalized over a O site and its two Cu neighbors, considered by Emery and Reiter \cite{ER88,ER90}.

Let us stress our assumption that spinon and holon bind not because of confinement but because of a short range attraction. Therefore, the bound state (\ref{F1}) can break up at an energy cost of order the antiferromagnetic exchange, and  the holon and spinon appear as gapped, free excitations which would contribute two-particle continuum spectra to photoemission or neutron scattering spectra. 
These fractionalized states can be included in our dimer model by expanding the Hilbert space to include monomers, but we will not do so here because we focus on the lowest energy sector. As a consequence, there is no monomer Fermi surface \cite{RK08} in the present model of the pseudogap metal.

The states (\ref{D1}) and (\ref{F1}) are precisely those that dominate in the 2-site dynamical mean field theory (DMFT) analysis of the Hubbard model by Ferrero {\em et al.} \cite{Ferrero09}: they correspond to the $S$ and $1+$ states of Ref.~\onlinecite{Ferrero09} respectively, which are shown in their Fig.~15 to be the dominant components of the ground state wavefunction at small $p$ (see also Ref.~\onlinecite{Tremblay11}). 
The DMFT analysis captures important aspects of pseudogap physics, but with a coarse momentum resolution of the Brillouin zone. 
In DMFT, the states on the 2-site cluster interact with a self-consistent environment in a mean-field way: the equations have so far
only been solved at moderate temperatures and the nature of the ultimate ground state at low doping remains unclear.
Our dimer model is a route to going beyond DMFT, and to include the non-trivial entanglement between these states on different pairs of sites
in a non-mean field manner. The local constraints between different pairs of dimers are accounted for, allowing for the emergence of gauge degrees
of freedom.

The original RK model can be mapped to a compact U(1) lattice gauge theory \cite{NRSS90,EFSK90,MVSS99}.
In the doped dimer models studied earlier, the monomers then carry U(1) gauge charges of $\pm 1$ on the two sublattices.
By the same reasoning, we see that the $F_{\eta\alpha}$ fermions carry no net gauge charge, but are instead
{\it dipoles\/} under the U(1) field. 

We can now describe our realization of the pseudogap metal. We envisage a state where the confinement length scale of the compact U(1) gauge field
is large, and specifically, larger than the spacing between the $F_{\eta \alpha}$ fermions. Then the $F_{\eta\alpha}$ fermions 
can move coherently in the presence of a dipolar coupling to the gauge fluctuations \cite{YQSS10}, and they will form Fermi surfaces
enclosing total area $p$, thus realizing a FL* state. The confinement  scale becomes large near the solvable RK point in the RK model \cite{AVLBTS04,Sondhi04},
near a Higgs transition to a $\mathbb{Z}_2$ spin liquid induced by allowing for diagonal dimers \cite{SSNR91,SSkagome,MS01,HYSK12},
or more generally near a deconfined critical point \cite{RK07}.
Our approach yields a `minimal model' for realizing FL* (which can be a stable, deconfined state in the $\mathbb{Z}_2$ spin liquid case),
and confinement transitions in metals. 

\begin{figure}
\begin{center}
\includegraphics[width=2.9in]{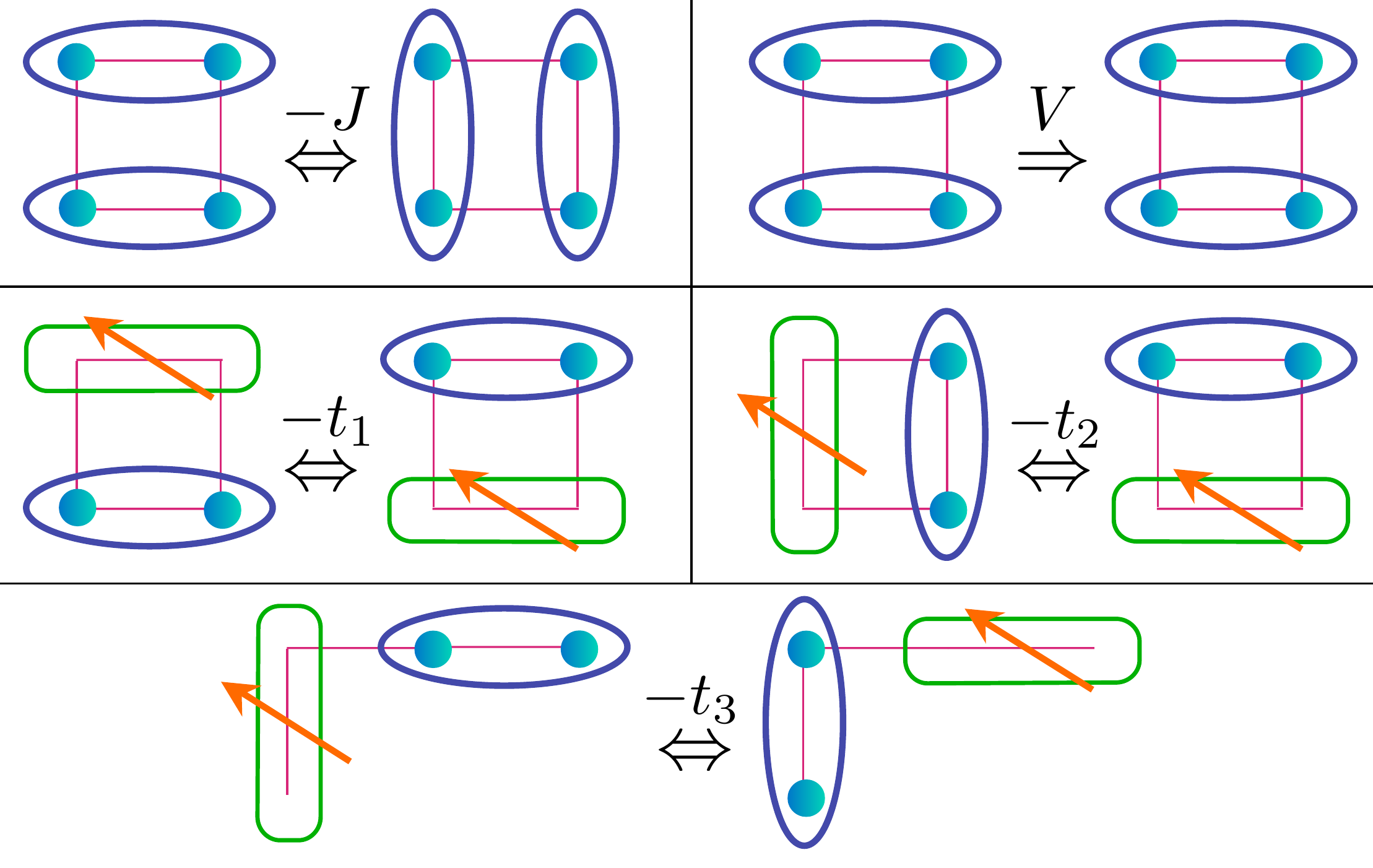}
\end{center}
\caption{(Color online) Terms in the Hamiltonian $H_{\rm RK} + H_1$.}
\label{fig:ham}
\end{figure}

We present results below for the following Hamiltonian, illustrated 
in Fig.~\ref{fig:ham}, acting on the dimer Hilbert space described above
\bea
H &=& H_{\rm RK} + H_1 + H_2 \nn
H_{\rm RK} &=& \sum_i \left[ -J \, D_{ix}^\dagger D_{i+\hat{y},x}^\dagger D_{iy}^{\vphantom\dagger} D_{i+\hat{x},y}^{\vphantom\dagger} + \mbox{~1 term} \right. \nn
&~&~~ \left. +V \, D_{ix}^\dagger D_{i+\hat{y},x}^\dagger D_{ix}^{\vphantom\dagger} D_{i+\hat{y},x}^{\vphantom\dagger} + \mbox{~1 term} \right] \nn
H_{1} &=& \sum_i \left[ -t_1 \, D_{ix}^\dagger F_{i+\hat{y},x \alpha}^\dagger F_{ix \alpha}^{\vphantom\dagger} D_{i+\hat{y},x}^{\vphantom\dagger} + \mbox{~3 terms} \right. \nn
&~&~~ \left. -t_2 \, D_{i+\hat{x},y}^\dagger F_{iy\alpha}^\dagger F_{ix\alpha}^{\vphantom\dagger} D_{i+\hat{y},x}^{\vphantom\dagger} + \mbox{~7 terms} \right. \nn
&~&~~ \left. -t_3 \, D_{i+\hat{x}+\hat{y},x}^\dagger F_{iy\alpha}^\dagger F_{i+\hat{x}+\hat{y},x\alpha}^{\vphantom\dagger} D_{iy}^{\vphantom\dagger} + \mbox{~7 terms} \right. \nn
&~&~~ \left. -t_3 \, D_{i+2\hat{y},x}^\dagger F_{iy\alpha}^\dagger F_{i+2\hat{y},x\alpha}^{\vphantom\dagger} D_{iy}^{\vphantom\dagger} + \mbox{~7 terms} \right], \label{ham}
\eea
where the undisplayed terms are generated by operations of the square lattice point group on the terms above. The first term, $H_{\rm RK}$, co-incides with the RK model for the undoped dimer model at $p=0$. Single fermion hopping terms are contained in $H_1$, with hoppings $t_i$ which are expected to be larger than $J$. A perturbative estimate of the dimer hopping amplitudes $t_i$ in terms of electron hopping parameters can be found in the SI Appendix. 
Note that all such terms must preserve the dimer close-packing constraint on every site, and 
we have chosen 3 terms with short-range hopping; longer-range hopping terms for the fermonic dimers are also possible, but expected to decay with distance
and are omitted for simplicity. Finally, $H_2$ allows for interactions between the fermionic dimers, with terms of the form
\beq
H_2 \sim \sum_i  (F_{ix\beta}^\dagger F_{i+\hat{y},x \alpha}^\dagger -  F^\dagger_{ix\alpha} F^\dagger_{i+y,x\beta}  ) F_{iy \beta}^{\vphantom\dagger} F_{i+\hat{x},y\alpha}^{\vphantom\dagger} + \ldots
\eeq
which preserve the dimer constraint and spin rotation invariance. Purely fermionic dimer models with similar dimer hopping terms have been considered by Pollmann \emph{et al.}~[\onlinecite{Pollmann2011}].

\section{Results}

We now present results for the dispersion and quasiparticle residue of a single fermion described by $H_{\rm RK} + H_1$; the interaction terms in $H_2$ play no role here. At a small $p$, the interactions between the fermionic dimers can be treated by a dilute gas expansion in $p$, while the dominant contributions to the quasiparticle dispersion and residue arise from the interaction between a single fermion and the close-packed sea of bosonic dimers. 
We computed the latter effects by exactly diagonalizing the singe fermion Hamiltonian on lattice sizes up to $8 \times 8$ with periodic boundary conditions, with the largest matrix of linear size 76,861,458. The RK model has two conserved winding numbers in a torus geometry, and these conservation laws also hold for our model: all results presented here are for the case of zero winding numbers. We extend these results to non-zero fermion density by interpolation in the SI Appendix.

\begin{figure}
\begin{center}
\includegraphics[width=2.7in]{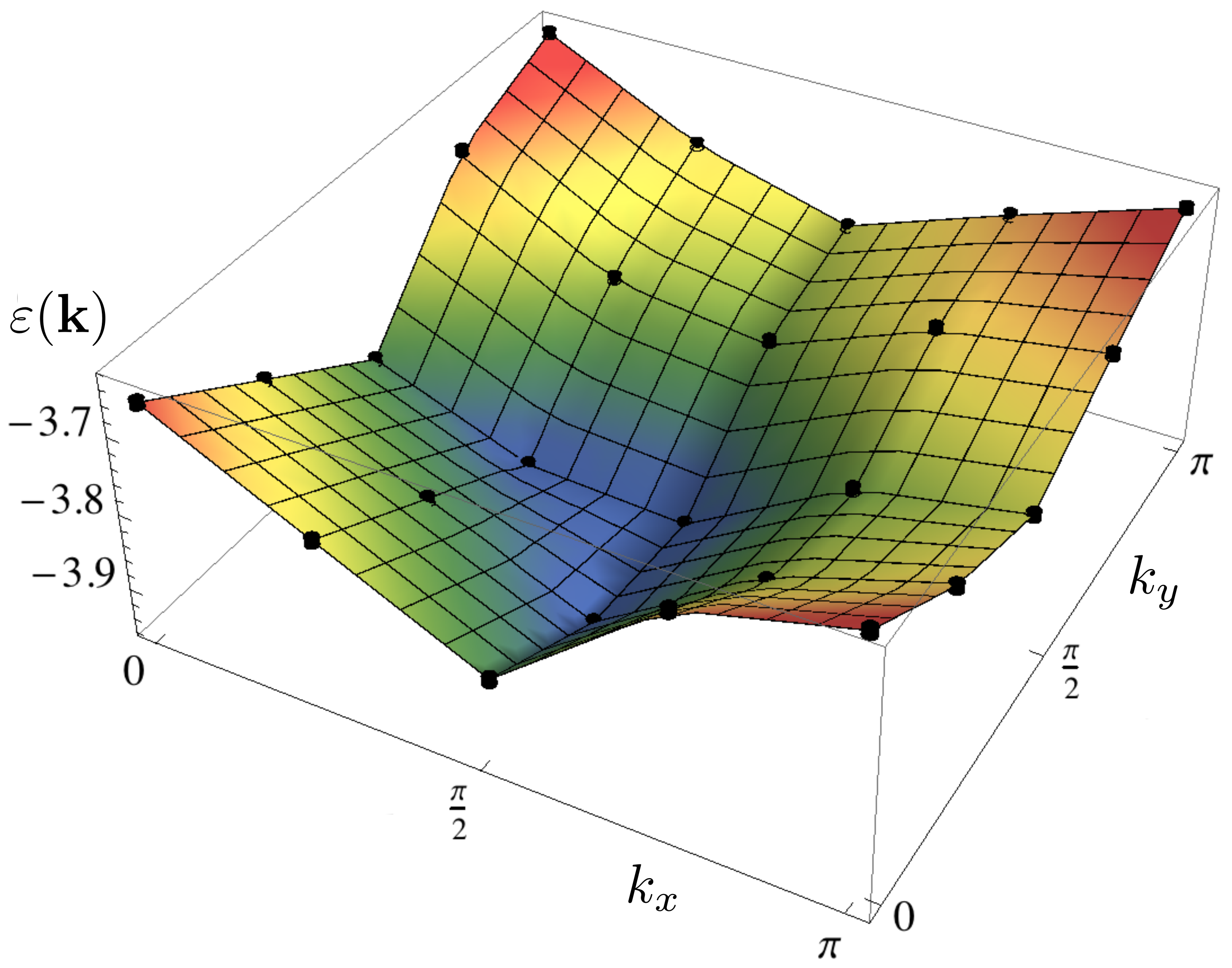}
\end{center}
\caption{(Color online) Lowest energy of a single charge $+e$ $F_{\eta \alpha}$ fermion as a function of momentum ${\bf k}$. We take hopping parameters obtained
from the $t$-$J$ model, 
$t_1 = -1.05$, $t_2 = 1.95$, and $t_3 = -0.6$, at the RK point $V=J=1$ on 
a $8\times 8$ lattice with periodic boundary conditions and zero winding numbers.
Note that the dispersion is not symmetric about the magnetic Brillouin zone boundary {\it i.e.\/} across the line
connecting $(\pi, 0)$ to $(0, \pi)$. Line cuts of this dispersion are in the SI Appendix.}
\label{fig:disp4}
\end{figure}

Our numerical study explored the dispersion of a single fermion over a range of values of the hopping parameters. We show in Fig.~\ref{fig:disp4} the dispersion $\varepsilon ({\bf k})$ for a single $F_{\eta \alpha}$ fermion for hopping parameters obtained by a perturbative connection on a $t$-$J$ model appropriate for the cuprates at the RK point $V=J=1$. The SI Appendix has similar results for additional parameter values. 

The minima of the fermion dispersion were found at different points in the Brillouin zone, but there was a wide regime with minima near momenta ${\bf k} = (\pm \pi/2, \pm \pi/2)$. In fact, for the momentum points allowed on a $8 \times 8$ lattice, the global minimum of the dispersion in Fig.~\ref{fig:disp4} is exactly at $(\pi/2, \pi/2)$. However, it is also clear from the figure that the dispersion is not symmetric about the antiferromagnetic Brillouin zone boundary, and that any interpolating function will actually have a minimum at $(k_m, k_m)$ with $k_m < \pi/2$. A dispersion with these properties is of experimental interest because it will lead to formation of hole pockets near the minima for the dimer model with a non-zero density of  $F_{\eta \alpha}$ fermions. Fig.~\ref{fig:cuts4} shows that changes to the dispersion from a $6 \times 6$ lattice are smaller than $5 \%$.

\begin{figure}
\begin{center}
\includegraphics[width=2.7in]{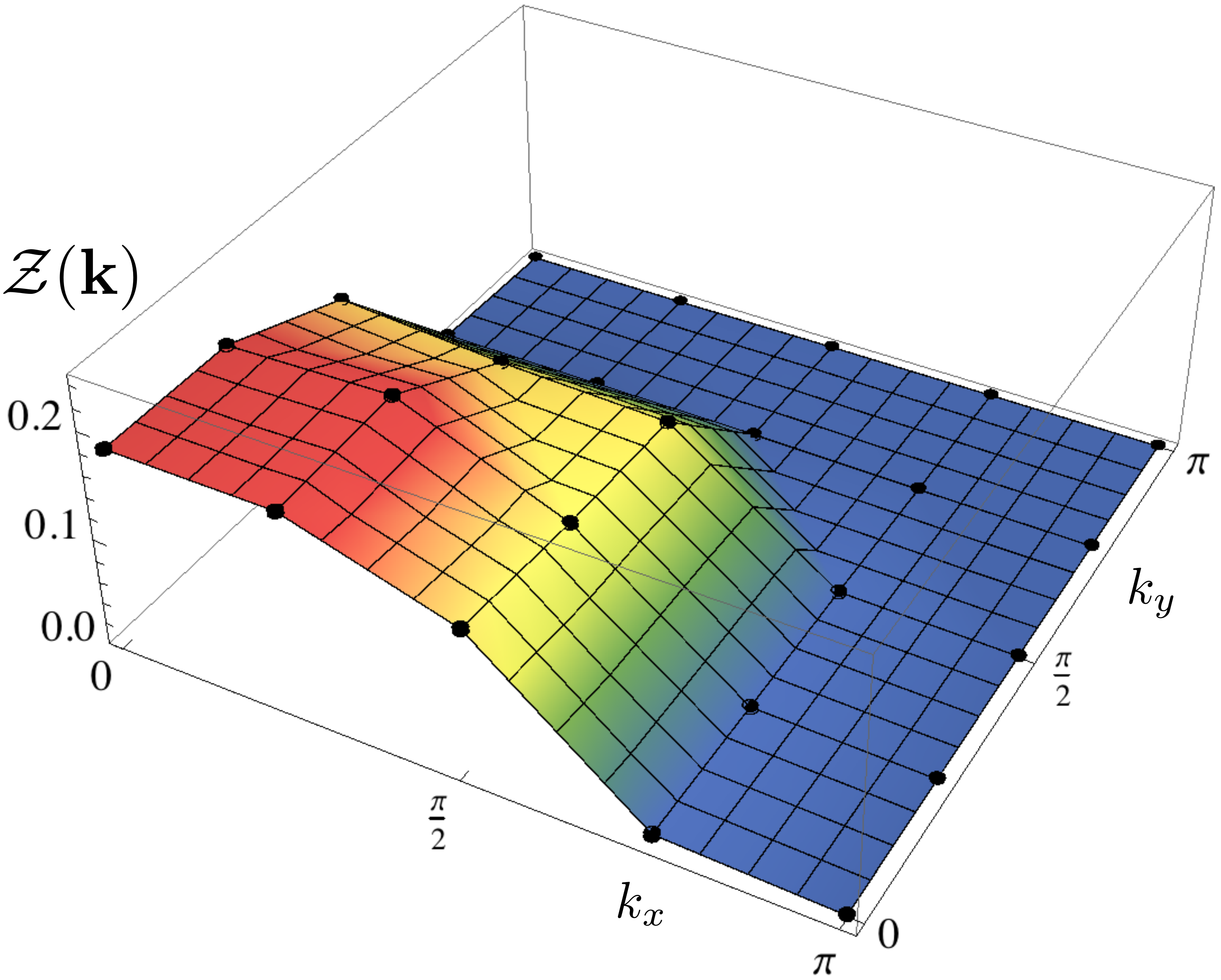}
\end{center}
\caption{(Color online) Quasiparticle residue of a charge $+e$ fermion computed from Eq.~(\ref{resdef}) for the parameters in Fig.~\ref{fig:disp4}, for a
 $8 \times 8$ lattice. The symmetry of the wavefunction yields
$\mathcal{Z}({\bf k}) = 0$ for all points between $(\pi, \pi)$ and $(\pi, 0)$.  Line cuts of $\mathcal{Z}({\bf k})$ are in the SI Appendix.}
\label{fig:res4}
\end{figure}
Our numerical results also yield interesting information on the quasiparticle residue of the electron operator. This is non-trivial even for the
case of a single fermionic dimer, because unlike a free electron, a fermionic dimer can only move by `resonating' with the background
of bosonic dimers, as is clear from Fig.~\ref{fig:ham}. In the presence of a finite density of fermionic dimers, there will be an additional renormalization
from the interaction between the fermions which we will not compute here. We don't expect this to have a significant ${\bf k}$ dependence around
the hole pockets. In the dimer model subspace defined by the states in Eqs.~(\ref{D1}) and (\ref{F1}), 
the electron annihilation operator on site $i$ has the same matrix elements as
\bea
C_{i \alpha}  &=& \frac{\epsilon_{\alpha\beta}}{2} \left( F^\dagger_{ix\beta} D_{ix}^{\vphantom\dagger} + F^\dagger_{i-\hat{x},x\beta} D_{i-\hat{x},x}^{\vphantom\dagger} \right. \nn
&~&~~~~~~~~~~~\left. + F^\dagger_{iy\beta} D_{iy}^{\vphantom\dagger} + F^\dagger_{i-\hat{y},y\beta} D_{i-\hat{y},y}^{\vphantom\dagger} \right),
\label{CFD}
\eea
relating the site to the 4 bonds around it ($\epsilon_{\alpha\beta}$ is the unit antisymmetric tensor).
Then the quasiparticle residue is obtained by computing
\beq
\mathcal{Z} ({\bf k})  = \left|\left\langle \Psi_F ({\bf k}) \right|   C_\alpha (-{\bf k}) \left| \Psi_{\rm RK} \right\rangle \right|^2 \,, \label{resdef}
\eeq
where $\left| \Psi_{\rm RK} \right\rangle $ is the ground state of the undoped model $H_{\rm RK}$, and $\left| \Psi_{F} ({\bf k}) \right\rangle$
is the ground state of $H_{\rm RK} + H_1 + H_2 $ in the sector with one $F_{\eta\alpha}$ fermion and total momentum ${\bf k}$
(the energy difference between these two states is $\varepsilon ({\bf k})$). We show the values of $\mathcal{Z} ({\bf k})$ in Fig.~\ref{fig:res4}, with
parameters the same as those in Fig.~\ref{fig:disp4}. 
Note the strong suppression of the residue in the second antiferromagnetic Brillouin zone; line-cut plots
of $\mathcal{Z} ({\bf k})$ in Fig.~\ref{fig:cuts4} highlight this suppression.
We found this suppression of $\mathcal{Z}({\bf k})$ to be a robust property in 
the regime of hopping parameters (with $t_2>0$) which had minima in the fermion dispersion along the Brillouin zone diagonal.
This result implies that the quasiparticle residue 
will be highly anisotropic around the hole pockets that appear in the finite fermion density 
case, with little spectral weight along the `back side' of the pocket. 

\begin{figure}
\begin{center}
\includegraphics[width=2.8in]{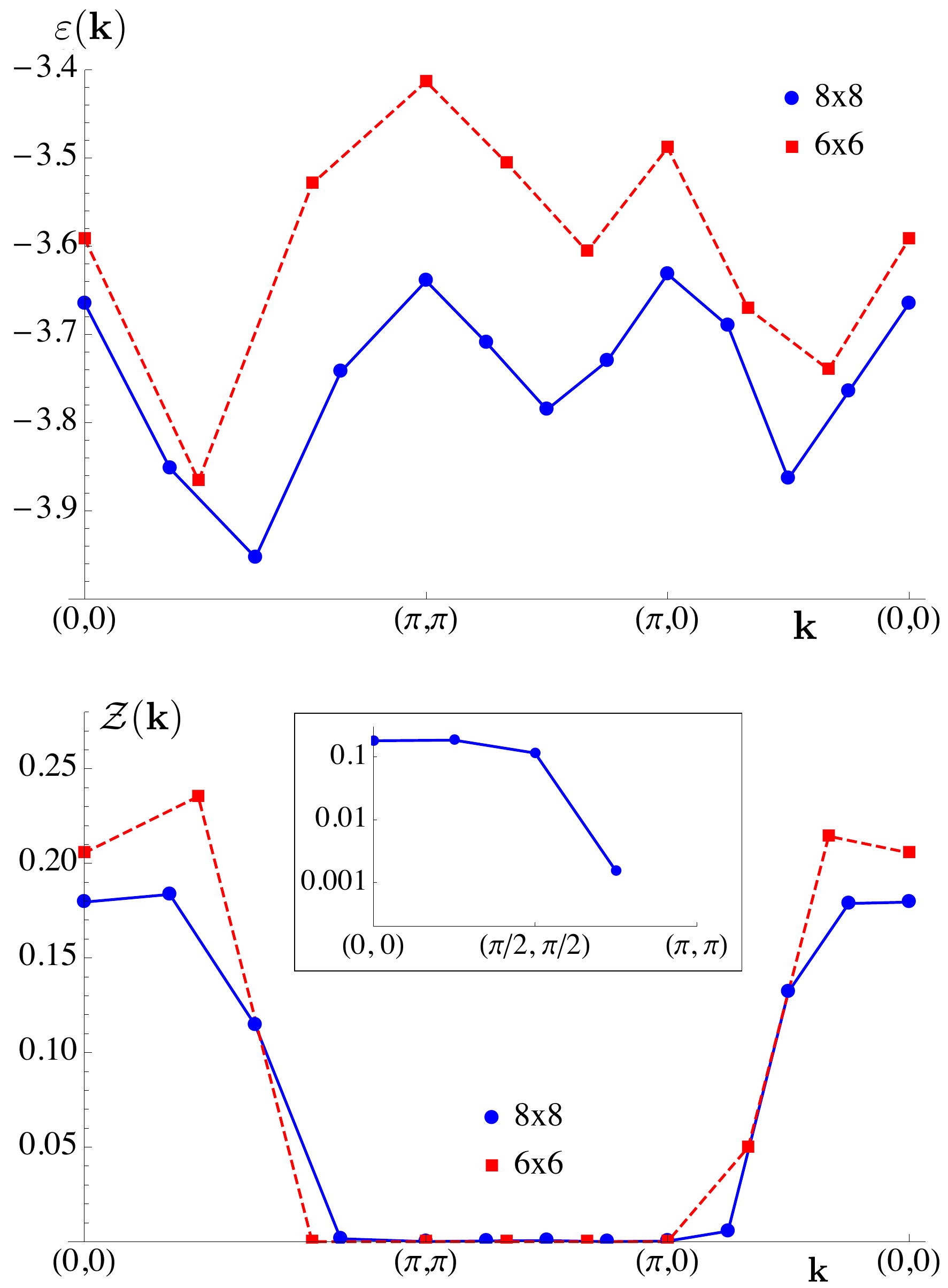}
\end{center}
\caption{Line cuts of the dispersion in Fig.~\ref{fig:disp4} (top), and of the quasiparticle residue in Fig.~\ref{fig:res4} (bottom).
Also shown are the results from exact diagonalization on a $6\times 6$ lattice for comparison (red squares), which has a different set of allowed momentum points. The overall shape of the dispersion remains the same as for the $8 \times 8$ lattice, and the fractional changes to $\varepsilon ({\bf k})$ are smaller than $5 \%$.
The inset shows the residue between $(0,0)$ and $(\pi, \pi)$ on a logarithmic scale. }
\label{fig:cuts4}
\end{figure}

It is also possible to study the system in perturbation theory in $t_i/J$. We begin with the model with one fermion at $t_1=t_2=t_3=0$. The problem reduces to that of finding the ground state of $H_{\rm RK}$ in the presence of a stationary fermionic dimer: it is possible to do this analytically at the solvable RK point $V=J$, as described in the SI Appendix. The fermion hoppings at non-zero $t_i$ is then computed perturbatively in a single-particle tight-binding model, with hopping matrix elements determined by overlaps of the wavefunctions with a stationary fermion. At the RK point, each matrix element reduces to the evaluation of a dimer correlation function in the classical problem of close-packed dimers on the square lattice \cite{Samuel80}. The computation of these matrix elements, and the resulting fermion dispersion is described in the SI Appendix. This perturbative dispersion is found to be in good agreement with our exact diagonalization results only for $|t_i /J| \lesssim 0.01$. This rather small upper limit is likely a consequence of the gaplessness of the RK point, 
so that higher order corrections involve non-integer powers of $t_i /J$.

\section{Discussion}

In this article we developed a new class of doped dimer models featuring coherent electronic quasiparticle excitations on top of a spin-liquid ground state. The scenario considered here is based on the assumption that spinons and holons form bound states on nearest neighbor sites. These fermionic bound states with spin $S=1/2$ and charge $+e$ form a Fermi sea with density $p$ and are observable as electronic quasiparticles in experiments. Such a Fermi sea realizes a topological quantum state called the `fractionalized Fermi liquid' \cite{TSSSMV03}, 
whose Fermi surfaces encloses an area distinct from the Luttinger value in a conventional Fermi liquid.

The undoped RK model on the square lattice features a deconfined spin liquid ground state only at the special RK point $J=V$, whereas the ground state breaks lattice symmetries away from this point. Consequently, our numerical results of a single fermionic dimer coupled to the background of bosonic dimers focused at the RK point to uncover properties of the FL* state. At finite densities of fermionic dimers we expect that our model \eqref{ham} features a FL* phase in an extended parameter range. However, similar to the RK model, we also expect a wide parameter regime where our model has a ground state with broken lattice symmetries. We leave the computation of the phase diagram of our model for future study.

The main implication of our model of the pseudogap metal (in zero applied magnetic field and at moderate $T$ below $T^\ast$) 
is that there are 4 well-formed pockets of charge $+e$ fermions carrying 
spin $S=1/2$ in the vicinity of (but not exactly centered at) momentum $(\pi/2, \pi/2)$. The total area enclosed by these pockets is $2 \pi^2 p$. Clearly, such pockets can immediately explain the Fermi liquid-like transport observed in recent optical \cite{Marel13}
and magnetoresistance \cite{MG14} measurements.
We also note that the hopping of electrons between CuO$_2$ layers requires to break either fermonic or bosonic dimers in our model, which naturally accounts for the observed gap in c-axis optical conductivity.

Experiments which involve removing one electron from the system (such as photoemission) have difficulty observing the `back sides'
of the pockets because of the small (but non-zero) quasiparticle residue $\mathcal{Z} ({\bf k})$ noted above (see Fig.~\ref{fig:res4}). 
We propose this feature as an explanation for the photoemission observation of Ref.~\onlinecite{PJ11} in the pseudogap metal.
For further studies of these pockets, it would be useful to employ
experimental probes of the Fermi surface which keep the electrons within the sample \cite{DCSS14c}: possibilities include ultrasound attenuation, 
optical Hall, and Friedel oscillations.

Our theory can be loosely summarized by `the electron becomes a dimer in the pseudogap metal', as in Eq.~(\ref{CFD}):
with a spin-liquid background present, there can be no single site state representing an electron, and a dimer is the simplest possibility.

The main advantage of our quantum dimer model over previous treatments \cite{RK07,RK08,YQSS10,LVVFL12,MPSS12} 
of fractionalized Fermi liquids (FL*) is that it properly captures lattice scale dispersions, quasiparticle residues, and Berry phases: all of these are expected to play crucial roles in the crossovers to confinement and associated symmetry breaking at 
low $T$ \cite{NRSS89,NRSS90,MVSS99}. Given the elongated dimer and dipolar nature
of the electron, Ising-nematic order \cite{KFE98} is a likely possibility; the 
$d$-form factor density wave \cite{SSRLP13,Fujita14} is then a plausible instability of such a nematic metal.
The interplay between the monopole-induced crossovers to 
confinement \cite{NRSS90,MVSS99}
and the density wave instabilities of the hole pockets \cite{DCSS14b,Mei14} can also be examined in such dimer models.
The onset of superconductivity will likely require additional states, such as a spinless, charge $+2e$ boson consisting of a pair of empty sites. 

\section{Acknowledgements}
  
We thank J.~Budich, D.~Chowdhury, J.~C.~Davis, D.~Drew, E.~Fradkin, A. Georges, S.~A.~Kivelson, A.~L\"auchli, A.~Millis, and E.~Sorensen for valuable discussions. K.~Fujita and J.~C.~Davis provided the phase diagram from Ref.~\cite{DFFDW15} which was
adapted to produce Fig.~\ref{fig:phasediag}.
We thank R.~Melko and D.~Hawthorn for CPU time at the University of Waterloo.
This research
was supported by the NSF under Grant DMR-1360789, the Templeton foundation, and MURI grant W911NF-14-1-0003 from ARO.
Research at Perimeter Institute is supported by the
Government of Canada through Industry Canada and by the Province of
Ontario through the Ministry of Research and Innovation. MP is supported by the ERC Synergy Grant UQUAM and SFB FOQUS of the Austrian Science Fund, as well as the Nano Initiative Munich (NIM).

\bibliography{cuprates}

\clearpage

\widetext
\vspace{5mm}
\begin{center}
\textbf{\large 
A quantum dimer model for the pseudogap metal\\
\vspace{0.05in}
Supplementary Information Appendix}\\~\\
Matthias Punk, Andrea Allais, and Subir Sachdev
\end{center}
\setcounter{equation}{0}
\setcounter{figure}{0}
\setcounter{table}{0}
\setcounter{page}{1}
\makeatletter
\renewcommand{\theequation}{S\arabic{equation}}
\renewcommand{\thesection}{S\Roman{section}}
\renewcommand{\thefigure}{S\arabic{figure}}

\newcommand{\dbar}{\mathrm{d}\hspace{-9pt}-\hspace{-3pt}}
\newcommand{\ii}{\mathrm{i}}
\newcommand{\dd}{\mathrm{d}}
\definecolor{ColorData_1_2}{rgb}{0.6,0.24,0.443}
\definecolor{ColorData_1_1}{rgb}{0.247,0.24,0.6}
\newcommand{\ket}[1]{\left|#1\right\rangle}
\newcommand{\bra}[1]{\left\langle#1\right|}
\newcommand{\expval}[1]{\left\langle#1\right\rangle}

\newcommand{\gridA}{
  \useasboundingbox (0.5ex, 0.5ex) rectangle (2.5 ex,2.5 ex);
  \draw [gray,very thin] (0.7ex, 1.0ex) -- (2.3ex, 1.0ex);
  \draw [gray,very thin] (0.7ex, 2.0ex) -- (2.3ex, 2.0ex);
  \draw [gray,very thin] (1.0ex, 0.7ex) -- (1.0ex, 2.3ex);
  \draw [gray,very thin] (2.0ex, 0.7ex) -- (2.0ex, 2.3ex);
}

\newcommand{\gridB}{
  \useasboundingbox (0.7ex, 0.5ex) rectangle (3.5 ex,2.5 ex);
  \draw [gray,thin] (0.7ex, 1.0ex) -- (3.3ex, 1.0ex);
  \draw [gray,thin] (0.7ex, 2.0ex) -- (3.3ex, 2.0ex);
  \draw [gray,thin] (1.0ex, 0.7ex) -- (1.0ex, 2.3ex);
  \draw [gray,thin] (2.0ex, 0.7ex) -- (2.0ex, 2.3ex);
  \draw [gray,thin] (3.0ex, 0.7ex) -- (3.0ex, 2.3ex);
}

\newcommand{\gridC}{
  \useasboundingbox (0.3ex, 0.5ex) rectangle (4.5 ex,4.5 ex);
  \draw [gray,thin] (0.3ex, 1.0ex) -- (4.3ex, 1.0ex);
  \draw [gray,thin] (0.3ex, 2.0ex) -- (4.3ex, 2.0ex);
  \draw [gray,thin] (0.3ex, 3.0ex) -- (4.3ex, 3.0ex);
  \draw [gray,thin] (0.3ex, 4.0ex) -- (4.3ex, 4.0ex);
  \draw [gray,thin] (1.0ex, 0.3ex) -- (1.0ex, 4.3ex);
  \draw [gray,thin] (2.0ex, 0.3ex) -- (2.0ex, 4.3ex);
  \draw [gray,thin] (3.0ex, 0.3ex) -- (3.0ex, 4.3ex);
  \draw [gray,thin] (4.0ex, 0.3ex) -- (4.0ex, 4.3ex);
}

\newcommand{\zerodimersA}{
\begin{tikzpicture}[baseline = 1.7ex]
  \gridC
\end{tikzpicture}}

\newcommand{\onedimerA}[1]{
\begin{tikzpicture}[baseline = 1.7ex]
  \gridC
  \draw [#1,very thick] (3.0ex, 2.0ex) -- (3.0ex, 3.0ex);
\end{tikzpicture}}

\newcommand{\onedimerB}[1]{
\begin{tikzpicture}[baseline = 1.7ex]
  \gridC
  \draw [#1,very thick] (2.0ex, 2.0ex) -- (2.0ex, 3.0ex);
\end{tikzpicture}}

\newcommand{\onedimerC}[1]{
\begin{tikzpicture}[baseline = 1.7ex]
  \gridC
  \draw [#1,very thick] (2.0ex, 3.0ex) -- (3.0ex, 3.0ex);
\end{tikzpicture}}

\newcommand{\onedimerD}[1]{
\begin{tikzpicture}[baseline = 1.7ex]
  \gridC
  \draw [#1,very thick] (1.0ex, 2.0ex) -- (2.0ex, 2.0ex);
\end{tikzpicture}}

\newcommand{\twodimersA}[2]{
\begin{tikzpicture}[scale=2,baseline = 2.2ex]
  \gridA
  \draw [#1,very thick] (1.0ex, 1.0ex) -- (1.0ex, 2.0ex);
  \draw [#2,very thick] (2.0ex, 1.0ex) -- (2.0ex, 2.0ex);
\end{tikzpicture}}

\newcommand{\twodimersB}[2]{
\begin{tikzpicture}[scale=2,baseline = 2.2ex]
  \gridA
  \draw [#1,very thick] (1.0ex, 1.0ex) -- (2.0ex, 1.0ex);
  \draw [#2,very thick] (1.0ex, 2.0ex) -- (2.0ex, 2.0ex);
\end{tikzpicture}}

\newcommand{\twodimersC}[2]{
\begin{tikzpicture}[scale=2,baseline = 2.2ex]
  \gridB
  \draw [#1,very thick] (1.0ex, 1.0ex) -- (2.0ex, 1.0ex);
  \draw [#2,very thick] (3.0ex, 1.0ex) -- (3.0ex, 2.0ex);
\end{tikzpicture}}

\newcommand{\twodimersD}[2]{
\begin{tikzpicture}[scale=1,baseline = 1.7ex]
  \gridC
  \draw [#1,very thick] (2.0ex, 2.0ex) -- (2.0ex, 3.0ex);
  \draw [#2,very thick] (3.0ex, 2.0ex) -- (3.0ex, 3.0ex);
\end{tikzpicture}}

\newcommand{\twodimersE}[2]{
\begin{tikzpicture}[scale=1,baseline = 1.7ex]
  \gridC
  \draw [#1,very thick] (3.0ex, 2.0ex) -- (3.0ex, 3.0ex);
  \draw [#2,very thick] (1.0ex, 2.0ex) -- (2.0ex, 2.0ex);
\end{tikzpicture}}

\newcommand{\manydimersA}[1]{
\begin{tikzpicture}[scale = 1,baseline = 1.7ex]
  \gridC
  \draw [#1,   very thick] (3.0ex, 2.0ex) -- (3.0ex, 3.0ex);
  \draw [black,very thick] (1.0ex, 1.0ex) -- (2.0ex, 1.0ex);
  \draw [black,very thick] (1.0ex, 2.0ex) -- (2.0ex, 2.0ex);
  \draw [black,very thick] (3.0ex, 1.0ex) -- (4.0ex, 1.0ex);
  \draw [black,very thick] (4.0ex, 2.0ex) -- (4.0ex, 3.0ex);
  \draw [black,very thick] (2.0ex, 4.0ex) -- (3.0ex, 4.0ex);
  \draw [black,very thick] (1.0ex, 3.0ex) -- (2.0ex, 3.0ex);
  \draw [black,very thick] (0.7ex, 4.0ex) -- (1.0ex, 4.0ex);
  \draw [black,very thick] (4.0ex, 4.0ex) -- (4.0ex, 4.5ex);
\end{tikzpicture}}

\newcommand{\manydimersB}[1]{
\begin{tikzpicture}[scale = 1,baseline = 1.7ex]
  \gridC
  \draw [#1,   very thick] (3.0ex, 2.0ex) -- (3.0ex, 3.0ex);
  \draw [black,very thick] (2.0ex, 2.0ex) -- (2.0ex, 3.0ex);
  \draw [black,very thick] (2.0ex, 1.0ex) -- (3.0ex, 1.0ex);
  \draw [black,very thick] (1.0ex, 1.0ex) -- (1.0ex, 2.0ex);
  \draw [black,very thick] (1.0ex, 4.0ex) -- (1.0ex, 4.5ex);
  \draw [black,very thick] (0.5ex, 3.0ex) -- (1.0ex, 3.0ex);
  \draw [black,very thick] (2.0ex, 4.0ex) -- (3.0ex, 4.0ex);
  \draw [black,very thick] (4.0ex, 3.0ex) -- (4.0ex, 4.0ex);
  \draw [black,very thick] (4.0ex, 1.0ex) -- (4.0ex, 2.0ex);
\end{tikzpicture}}
\definecolor{color1}{rgb}{0.8,0.3,0.3}

\newcommand{\elemA}{\mathord{\vcenter{\hbox{\includegraphics[height=2.5ex]{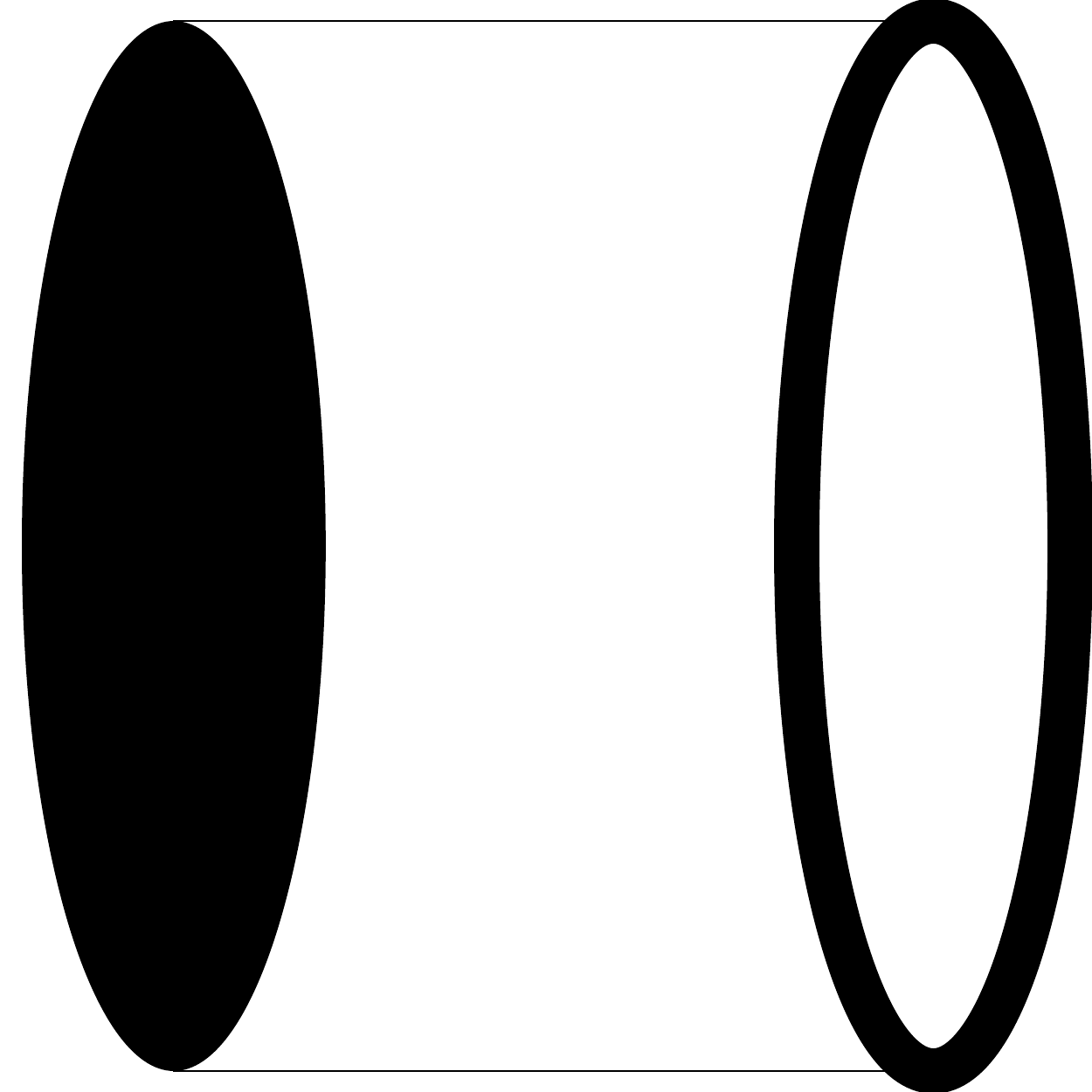}}}}}
\newcommand{\elemB}{\mathord{\vcenter{\hbox{\includegraphics[height=2.5ex]{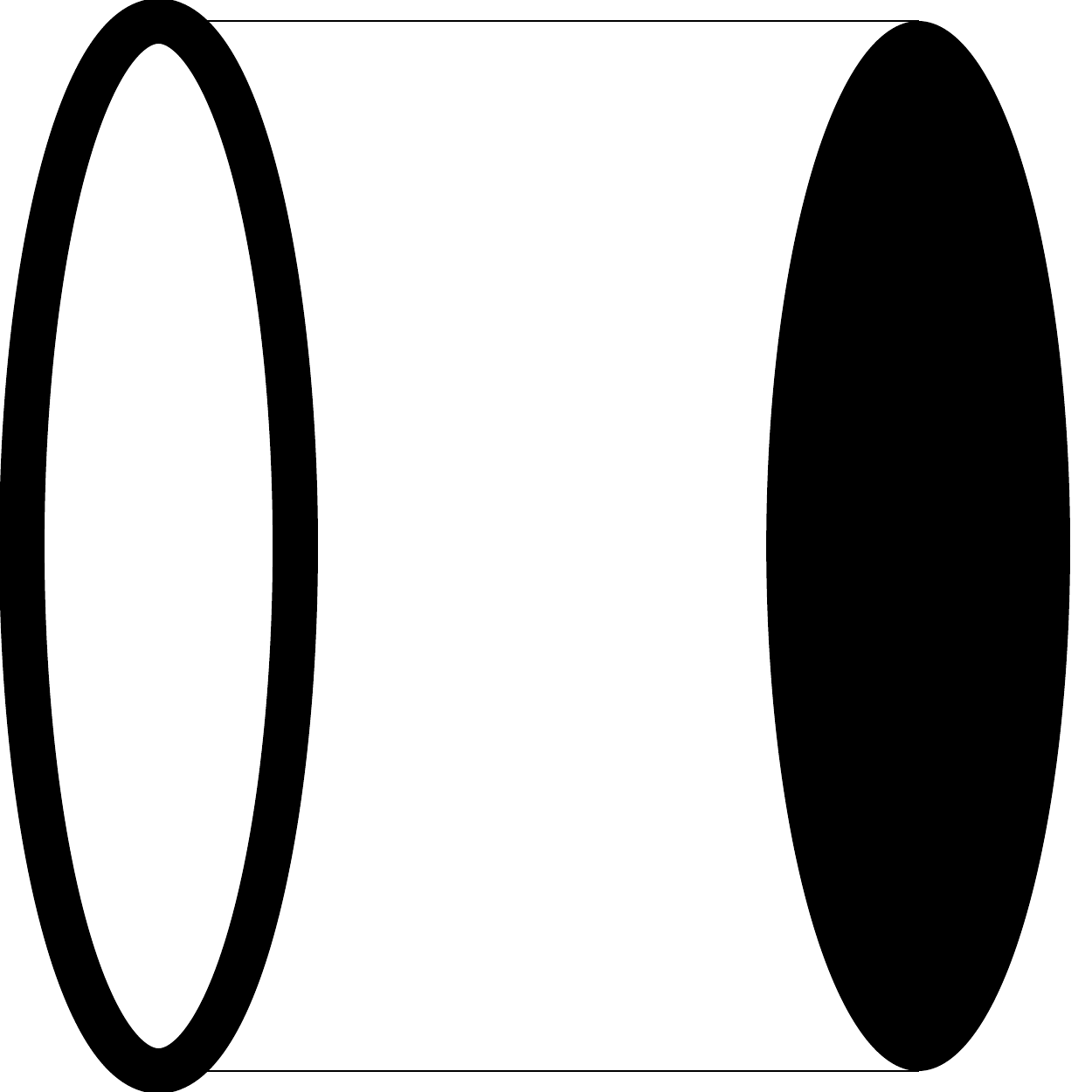}}}}}
\newcommand{\elemC}{\mathord{\vcenter{\hbox{\includegraphics[height=2.5ex]{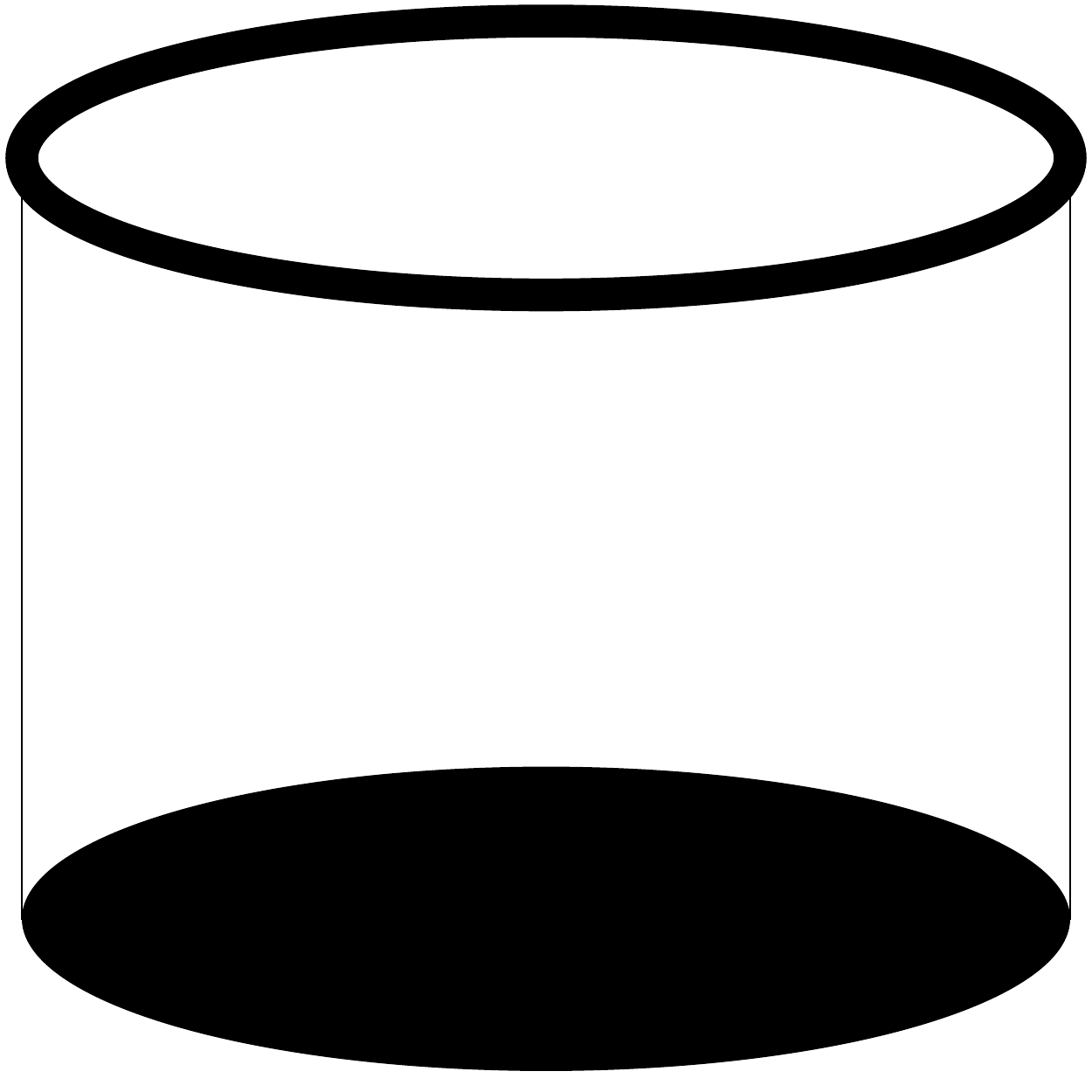}}}}}
\newcommand{\elemD}{\mathord{\vcenter{\hbox{\includegraphics[height=2.5ex]{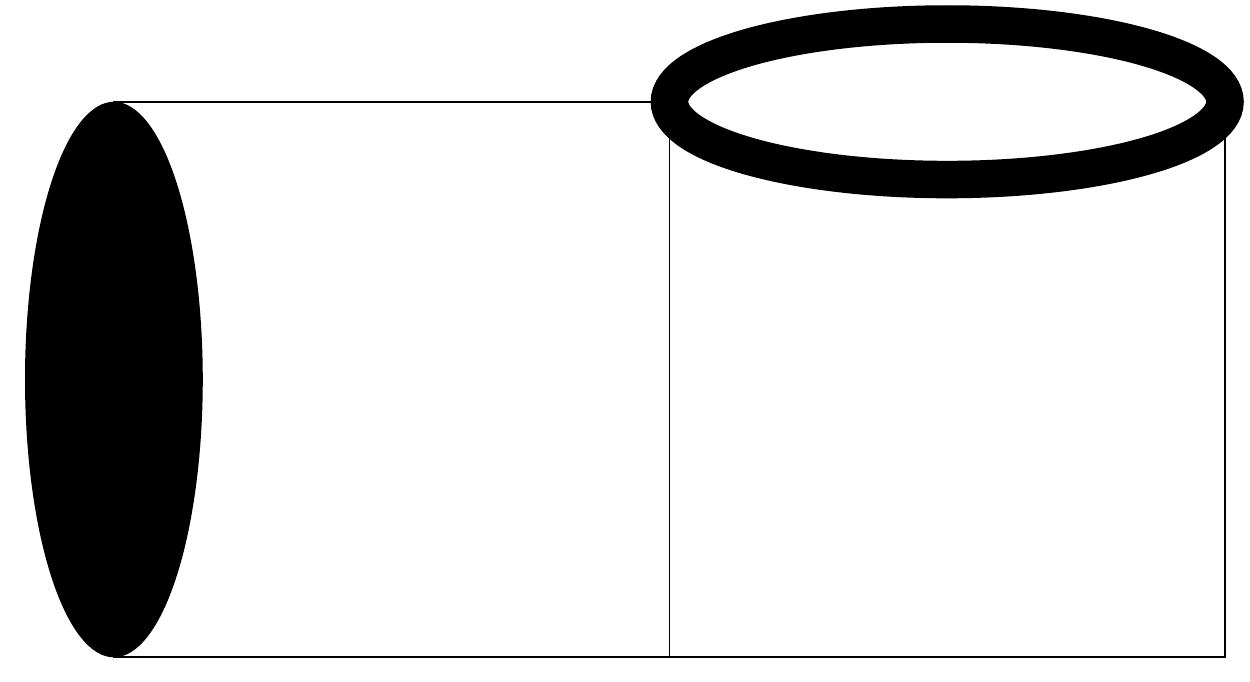}}}}}
\newcommand{\elemE}{\mathord{\vcenter{\hbox{\includegraphics[height=2.5ex]{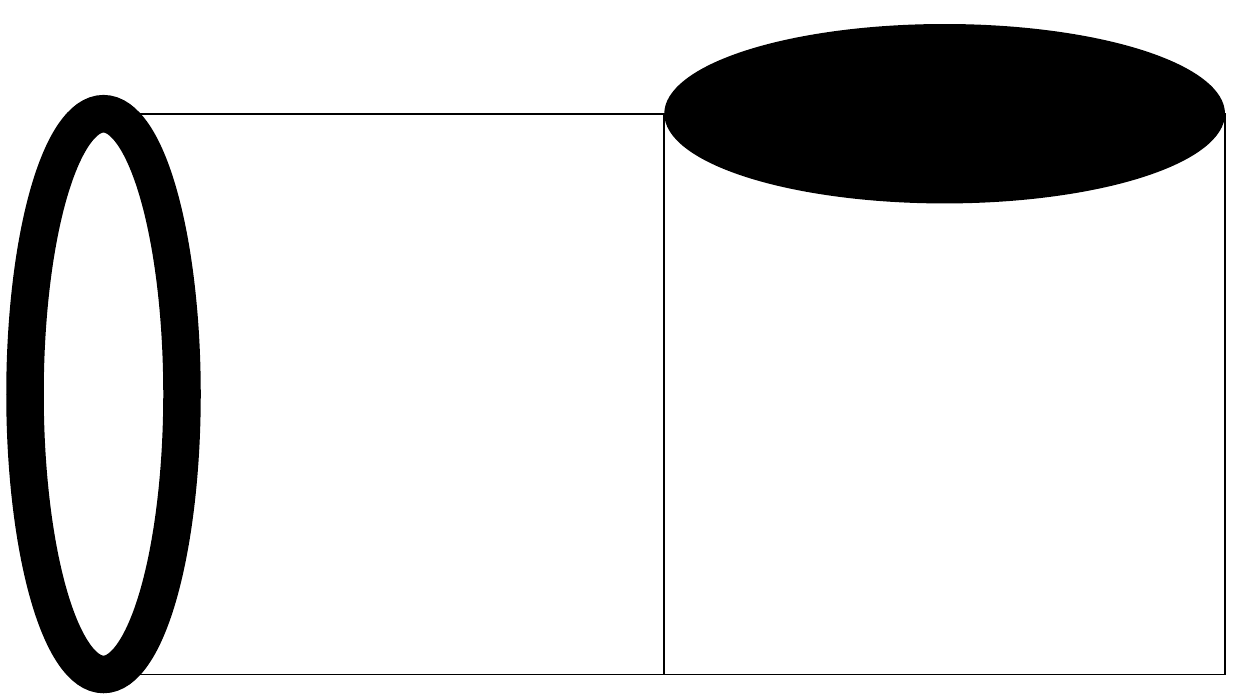}}}}}

\section{Connection to the $t$-$J$ model}
\label{sec:tJ}

A quantitive connection between the dimer model Hamiltonian and the $t$-$J$ model can only
be made in the limiting regime of a certain large $N$ SU($N$) antiferromagnet \cite{RS89}; we also require $t \ll J$ to allow
restriction to the dimer subspace in the doped case. Eventually, we hope that it will be possible to use {\em e.g.\/} DMFT methods \cite{Ferrero09}
to make this connection. Nevertheless, it is useful, for now, to determine the relationship between the hopping parameters of
the dimer and $t$-$J$ model in the perturbative approach of large $N$ and small $t$. We will directly work with the physical SU(2) case here.

The dimer hopping amplitudes are given by the matrix elements
 \begin{eqnarray}
 t_1 &=& - \langle \elemA | H_0 | \elemB \rangle \nn
 t_2 &=& - \langle \elemB | H_0 | \elemC \rangle \nn
 t_3 &=& - \langle \elemD | H_0 | \elemE \rangle \ ,
 \end{eqnarray}
where the fermonic (bosonic) dimer is represented by the black (open) ellipse and $H_0$ denotes the standard tight-binding hamiltonian of electrons including up to third-nearest neighbour hopping 
\begin{equation}
 H_0 = - t \sum_{\langle i j \rangle} \left(c^\dagger_{i \alpha} c^{\ }_{j\alpha} + \text{h.c.} \right) - t' \sum_{\langle \langle i j \rangle \rangle}\left(c^\dagger_{i \alpha} c^{\ }_{j\alpha} + \text{h.c.} \right)  - t'' \sum_{\langle \langle \langle i j \rangle \rangle \rangle} \left(c^\dagger_{i \alpha} c^{\ }_{j\alpha} + \text{h.c.} \right) .
 \end{equation}
Using the representation in terms of the dimer operators
 \begin{eqnarray}
| \elemA \rangle &=& D^\dagger_{i+\hat{x},y} \,  F^\dagger_{i y\uparrow} |0\rangle  \nn
 | \elemB \rangle &=& D^\dagger_{i y} \,  F^\dagger_{i+\hat{x},y\uparrow} |0\rangle \nn
 | \elemC \rangle &=& D^\dagger_{i+\hat{y},x} \,  F^\dagger_{i x\uparrow} |0\rangle \nn
 | \elemD \rangle &=& D^\dagger_{i+\hat{x}+\hat{y},x} \,  F^\dagger_{i y \uparrow} |0\rangle \nn
 | \elemE \rangle &=& D^\dagger_{i y} \,  F^\dagger_{i+\hat{x}+\hat{y}, x\uparrow} |0\rangle \ ,
  \end{eqnarray}
  we can obtain expressions in terms of the electron operators via the mappings in Eqs.~(\ref{D1}) and (\ref{F1}).
 Evaluating the matrix elements is now straightforward, and we obtain
\begin{eqnarray}
t_1 &=& - \frac{1}{2} \left( t+t' \right) \nn
t_2 &=& \frac{1}{2} \left( t-t' \right) \nn
t_3 &=& - \frac{1}{4} \left( t+t'+t'' \right) \ , \label{tvals}
\end{eqnarray}
where we used the mappings for the dimer operators in Eqs.~(\ref{D1}) and (\ref{F1}). 
Choosing electron hopping parameters suitable for cuprates, $t'/t =-0.3$, $t''/t=0.1$, we obtain
\begin{eqnarray}
t_1 &=& - 0.35 \, t \nn
t_2 &=& 0.65 \, t \nn
t_3 &=& - 0.2 \, t \ . \label{t123t}
\end{eqnarray}
The dispersion and residue with the hoppings obeying Eq.~(\ref{t123t}) with $t=3$
were shown in Figs.~\ref{fig:disp4} and~\ref{fig:res4} in the main text, and their line cuts appear
in Fig.~\ref{fig:cuts4}.

\section{Exact diagonalization}

This section contains additional plots of the energy eigenvalues and quasiparticle residues obtained by exact diagonalization of $H_{\rm RK}+ H_1$.
We note that the exactly solvable RK point of the undoped mode at $V=J=1$ can be extended to the doped case {\em e.g.\/} by including a nearest-neighbor
repulsion $V_1 = t_1$ between the $F_{\eta\alpha}$ and $D_{\eta}$ dimers, and setting $t_2=t_3=0$. However, the exact solution only
yields the energy at zero momentum, and numerics are required for the full dispersion even at this point.

\begin{figure}
 \begin{center}
 \includegraphics[width=6.8in]{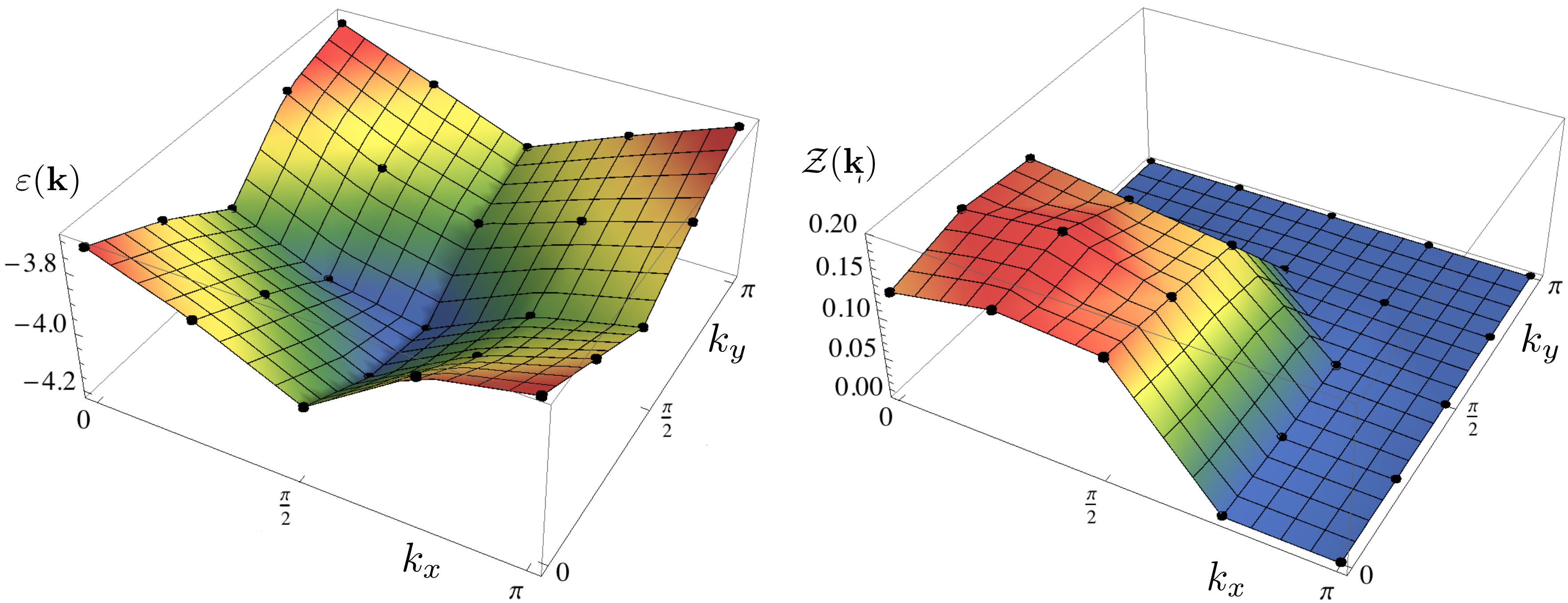}
 \caption{Dispersion and quasiparticles residue for $J=V=1$, and $t_1= t_2= -t_3= 1$
 on a $8 \times 8$ lattice.}
  \label{fig:disp1a}
 \end{center}
 \end{figure}

\begin{figure}
 \begin{center}
 \includegraphics[width=6.8in]{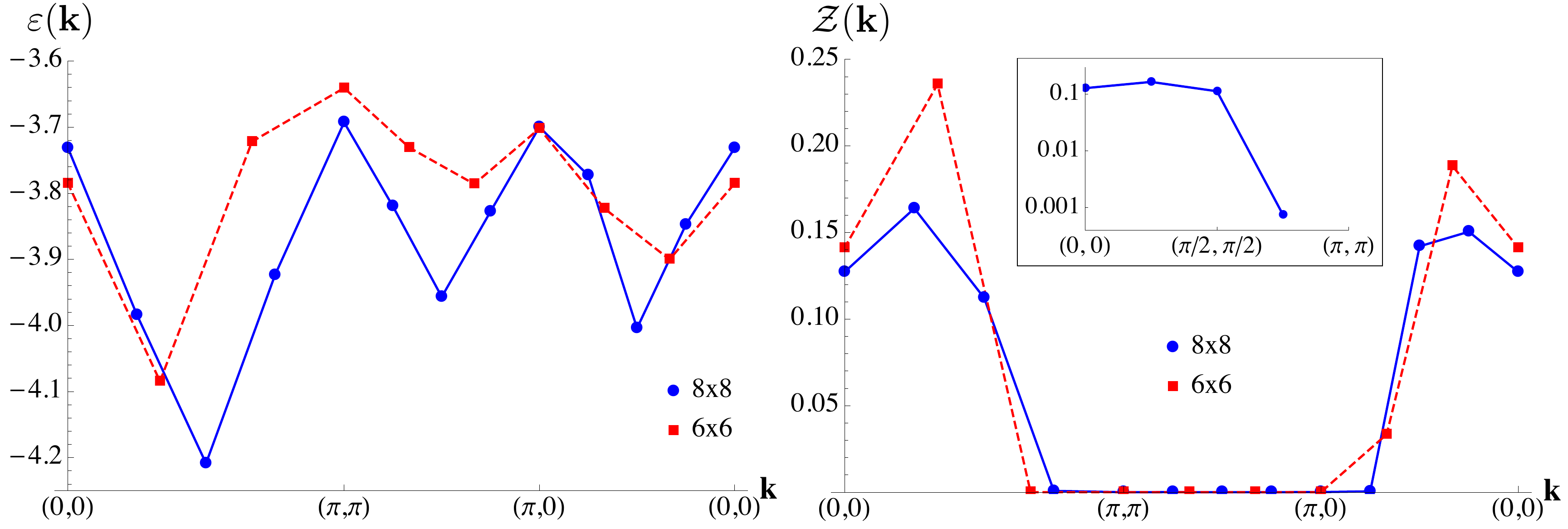}
 \caption{Line cuts of the dispersion and quasiparticles residue in Fig.~\ref{fig:disp1a}, along with corresponding data
 for a $6 \times 6$ lattice. }
  \label{fig:cuts1a}
 \end{center}
 \end{figure}

\begin{figure}
\begin{center}
\includegraphics[width=6.8in]{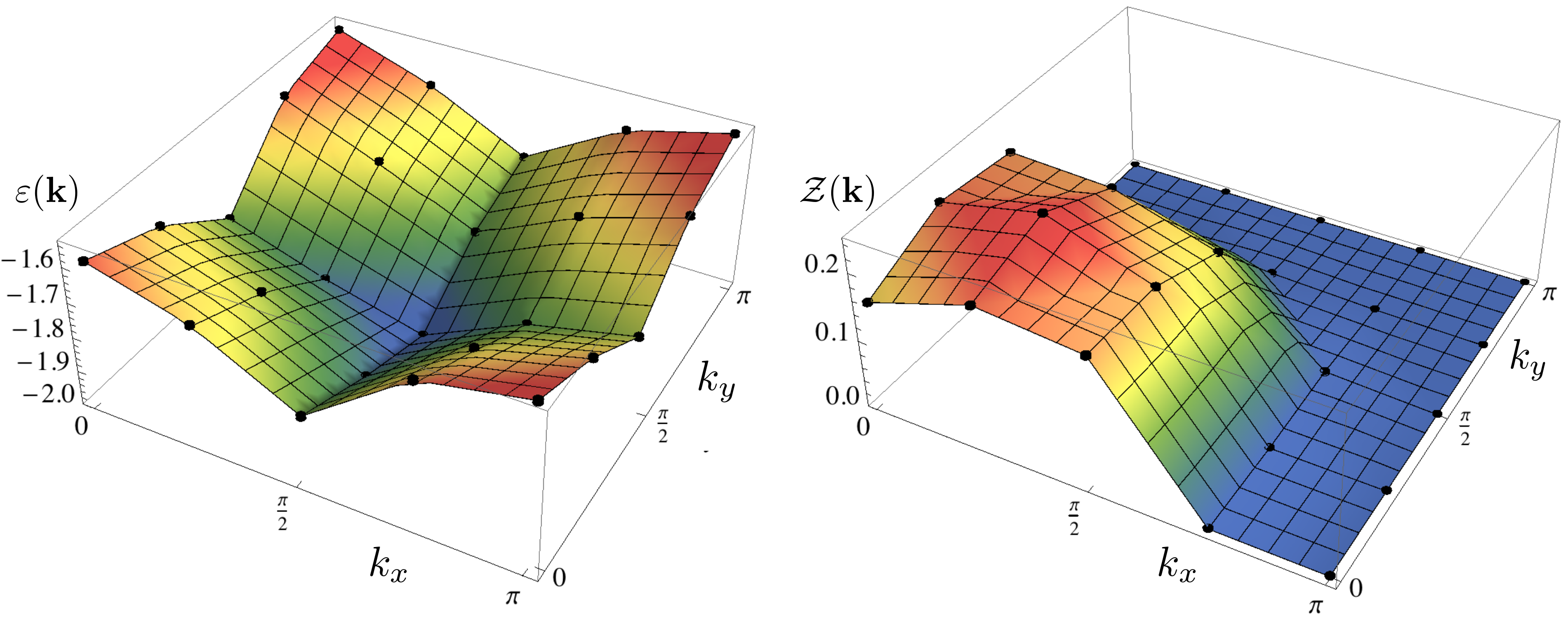}
\end{center}
\caption{As in Fig.~\ref{fig:disp1a}, with the parameters
 $t_1 = t_2 = - t_3 = 0.5$ at the RK point $V=J=1$.}
\label{fig:disp3}
\end{figure}

\begin{figure}
\begin{center}
\includegraphics[width=6.8in]{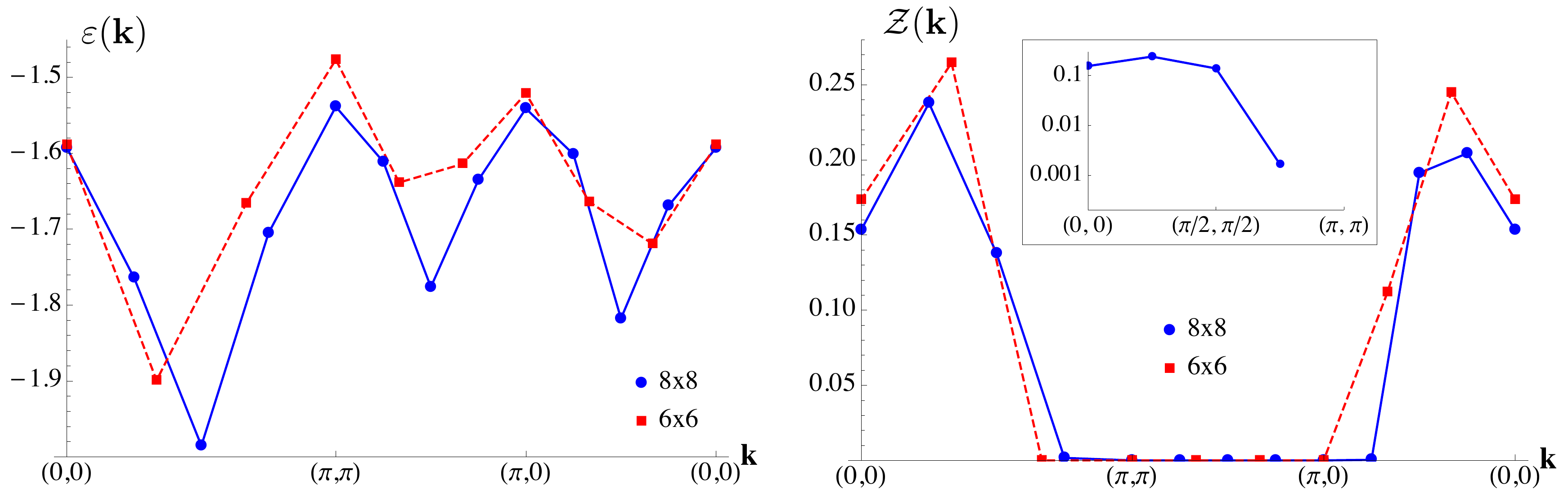}
\end{center}
\caption{As in Fig.~\ref{fig:cuts1a}, with the parameters
 $t_1 = t_2 = - t_3 = 0.5$ at the RK point $V=J=1$.}
\label{fig:cuts3}
\end{figure}

\begin{figure}
\begin{center}
\includegraphics[width=6.8in]{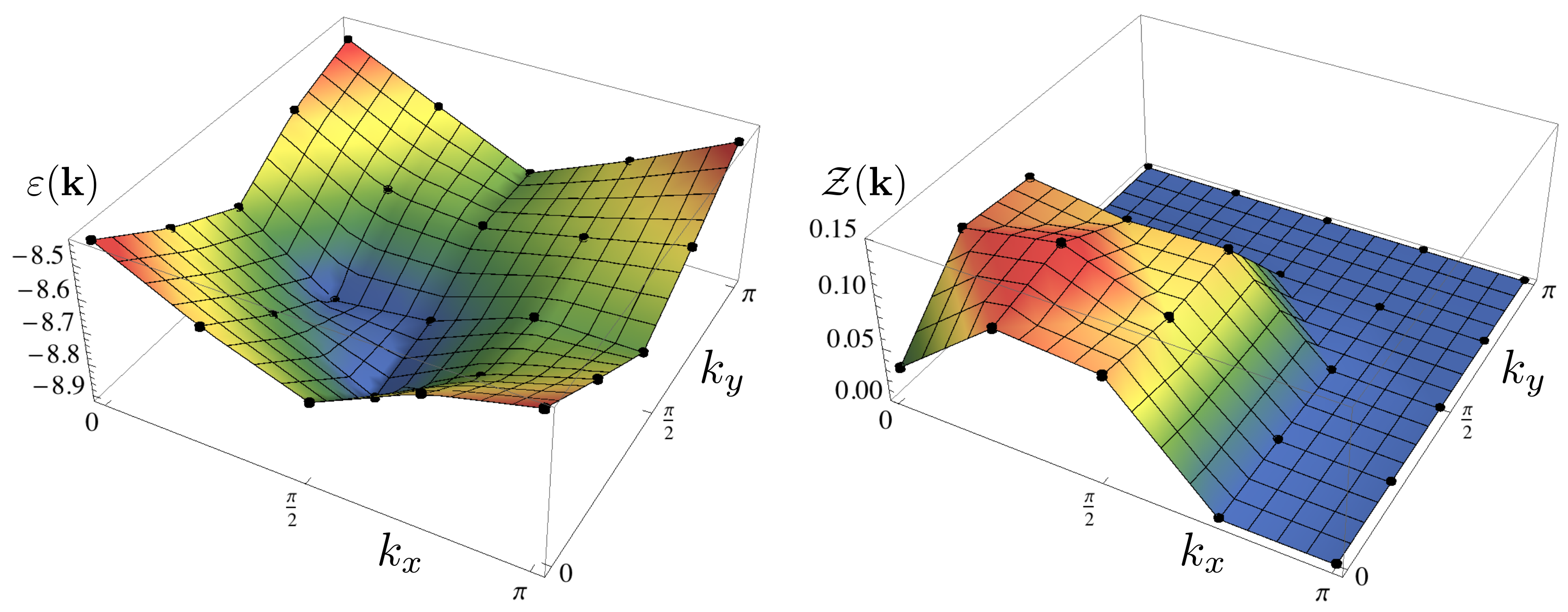}
\end{center}
\caption{As in Fig.~\ref{fig:disp1a}, with the parameters
 $t_1 = t_2 = - t_3 = 2$ at the RK point $V=J=1$.}
\label{fig:disp2}
\end{figure}

\begin{figure}
\begin{center}
\includegraphics[width=6.8in]{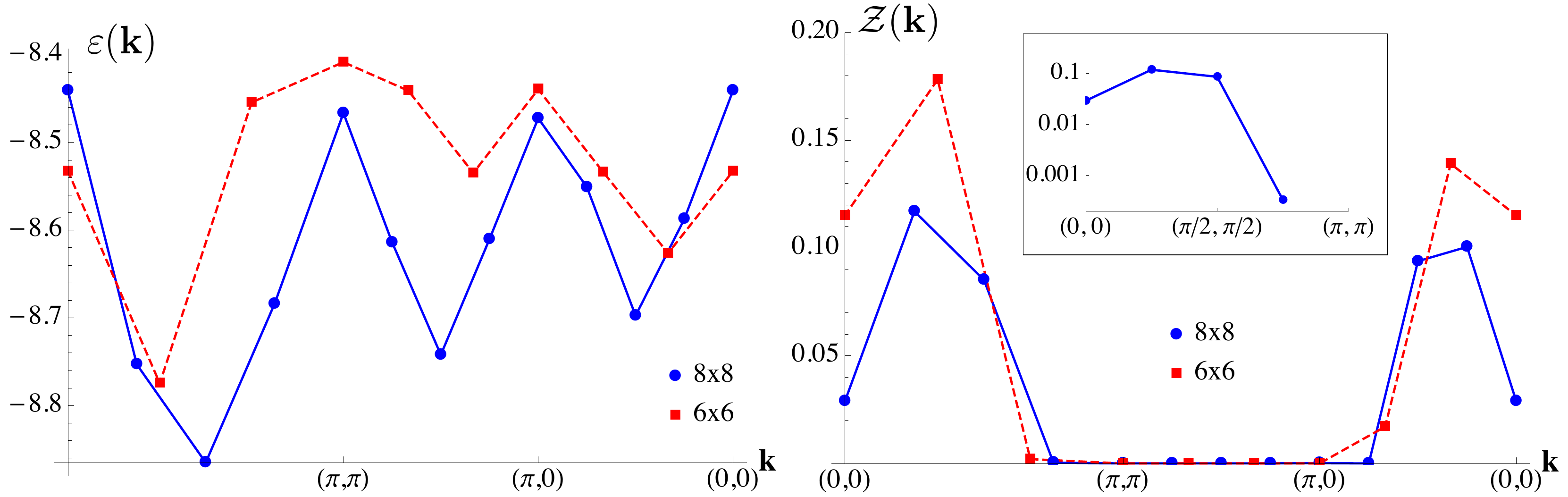}
\end{center}
\caption{As in Fig.~\ref{fig:cuts1a}, with the parameters
 $t_1 = t_2 = - t_3 = 2$ at the RK point $V=J=1$.}
\label{fig:cuts2}
\end{figure}

In Figs.~\ref{fig:disp1a} and~\ref{fig:cuts1a} we show the results for $J=V=1$ and $t_1=t_2=-t_3=1$.
We also consider the effect of uniformly scaling all the $t_i$'s while keeping $J$ fixed. 
Figs.~\ref{fig:disp3} and~\ref{fig:cuts3} contain results for $t_1=t_2=-t_3=0.5$, while Figs.~\ref{fig:disp2} and~\ref{fig:cuts2} contain results for $t_1=t_2=-t_3=2.0$.
Note that the while the general forms of the dispersion and quasiparticle residue remain the same when we uniformly scale the $t_i$,
the finite size corrections become larger at larger $t_i/J$.

\begin{figure}
 \begin{center}
\includegraphics[width=6.8in]{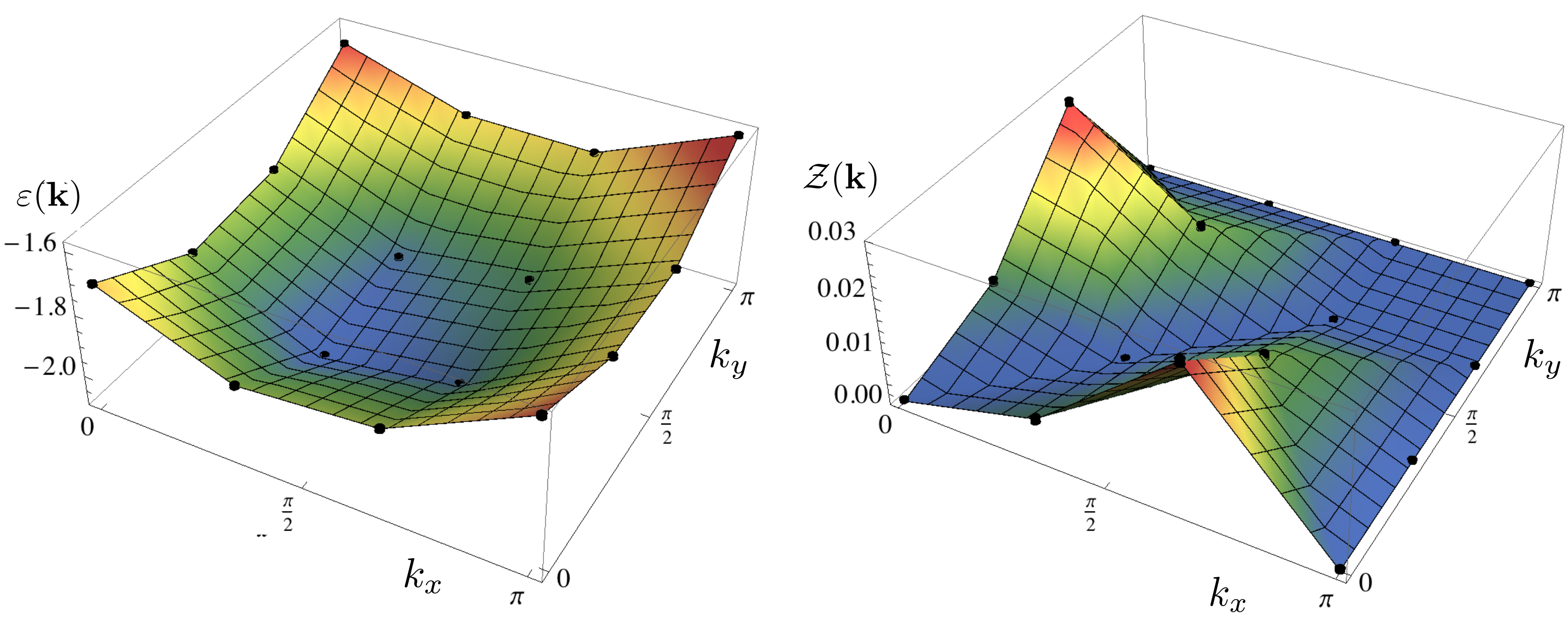}
 \caption{Dispersion and quasiparticles residue for $J=V=1$, and hopping parameters $t_1=-0.5$, $t_2=-1$ and $t_3=0.5$
 on a $6 \times 6$ lattice. For this case, the quasiparticle residue $\mathcal{Z} ({\bf k})$ vanishes for ${\bf k}$ along the diagonal 
 because the wavefunction is odd under reflections about the diagonal.}
  \label{fig:disp5}
 \end{center}
 \end{figure}

Lastly, in Fig.~\ref{fig:disp5}, we consider a set of parameters for which the minimum of the dispersion, $\varepsilon ( {\bf k})$ remains on the diagonal near
$(\pi/2, \pi/2)$, but the quasiparticle residue $\mathcal{Z} ({\bf k})$ vanishes for ${\bf k}$ along the diagonal. This happens because the wavefunction
of the $F_{x\alpha}$ fermions has the opposite sign of the wavefunction for the $F_{y \alpha}$ fermions; then by Eq. (\ref{CFD}), $\mathcal{Z} ({\bf k})$
will vanish for ${\bf k}$ along the diagonal.

\begin{figure}
 \begin{center}
\includegraphics[width=2.4in]{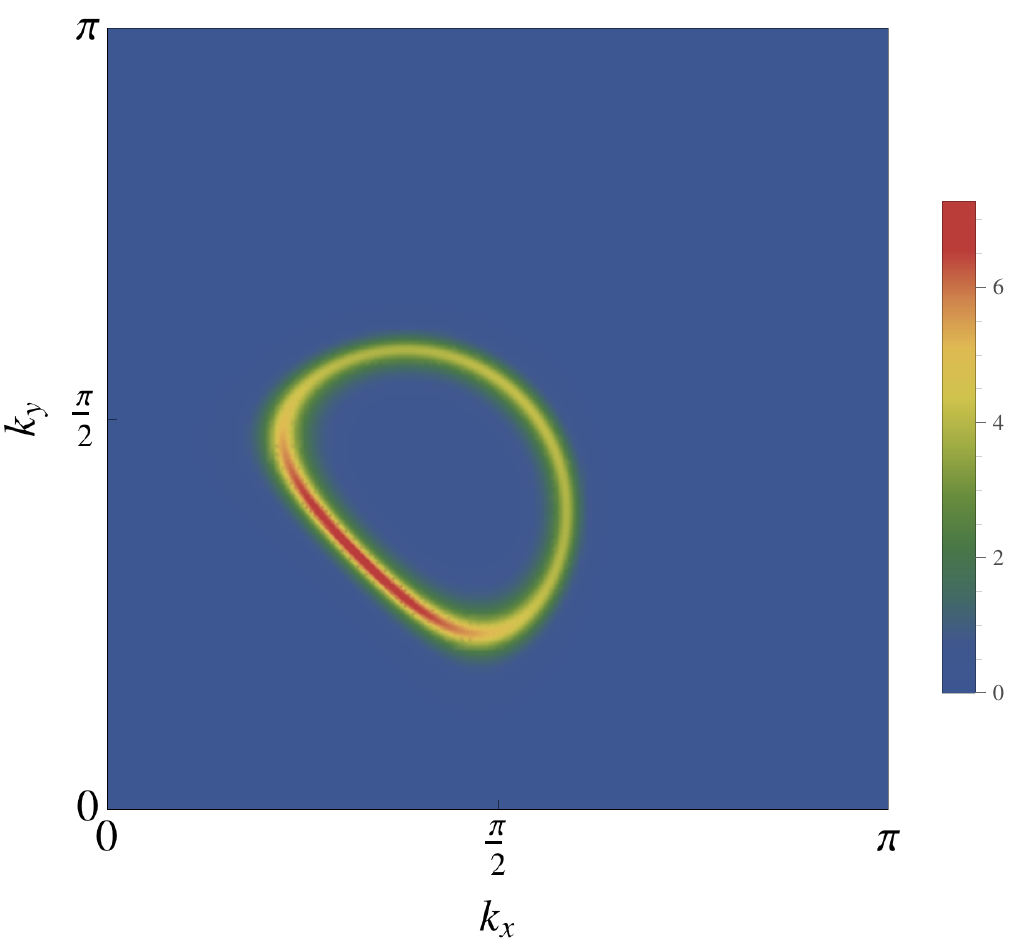}~~~~~~
\includegraphics[width=2.4in]{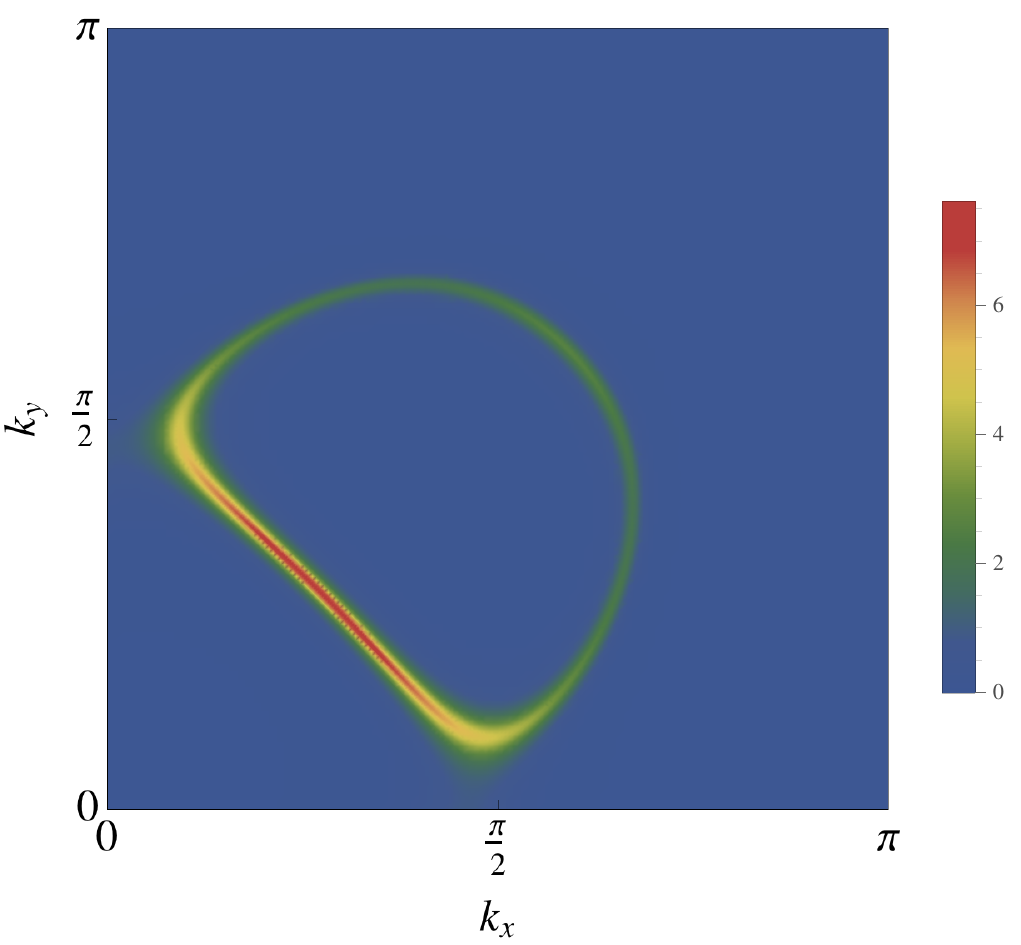}
 \caption{Electronic spectral functions $A({\bf k})=\mathcal{Z}({\bf k}) \, \delta(\varepsilon({\bf k})-\mu)$ at the Fermi level obtained by interpolating the data in Figs.~\ref{fig:disp4} and~\ref{fig:res4}. For illustrative purposes, we replace the delta function by a Lorentzian with finite width.
 Chemical potentials $\mu$ are chosen to obtain doping densities of $p = 0.083$ (left) and $0.225$ (right).}
  \label{fig:interp4}
 \end{center}
 \end{figure}

As an alternative way of presenting the above data, we used the information on the dispersion, $\varepsilon ( {\bf k})$, from Fig.~\ref{fig:disp4}, 
and the quasiparticle residue, $\mathcal{Z} ({\bf k})$, from Fig.~\ref{fig:res4}, and interpolated to all ${\bf k}$ in the Brillouin zone. The resulting electronic
spectral weights at the Fermi energy are shown in Fig.~\ref{fig:interp4} at two different doping densities.

\clearpage
\section{Perturbative computation}

This supplement will describe the perturbative evaluation of the single fermion dispersion for small $t_i$ under the Hamiltonian
$H_{\rm RK}+ H_1$ in Eq.~(\ref{ham}) at the RK point $V=J$.

At zeroth order in the $t_i$, we need the ground state of a stationary fermion. 
This ground state is given by an equal weight superposition of all dimer configurations with the fermionic dimer in a fixed position:
\begin{equation}
 \ket{\onedimerA{color1}} \equiv \frac{1}{\sqrt{N}}\left[\ket{\manydimersA{color1}} + \ket{\manydimersB{color1}} + \cdots\right],
\end{equation} 
where the fermionic dimer is represented by the red bar. The position of the fermionic dimer can be arbitrary, and so there
is a large degeneracy in the ground state subspace.

Let us denote the fermionic hopping terms $T_1$, $T_2$, $T_3$ with matrix elements $t_1$, $t_2$, $t_3$ respectively.
These operators $T_1, T_2, T_3$ lift the degeneracy of the ground state subspace. Within this subspace, the only non-zero matrix elements which are inequivalent under translations and rotations are:
\begin{align}
 &Z_1 \equiv \bra{\onedimerB{color1}}T_1\ket{\onedimerA{color1}}\,,
 &&Z_2 \equiv \bra{\onedimerC{color1}}T_2\ket{\onedimerA{color1}}\,,
 &&Z_3 \equiv \bra{\onedimerD{color1}}T_3\ket{\onedimerA{color1}}.
\end{align} 

If we introduce a partition function $Q$ that counts the number of single-flavor dimer close packing configurations with a certain number of dimers fixed, we have
\begin{align}
&Z_1 = Z_2 = \frac{Q\left[\twodimersD{black}{black}\right]}{Q\left[\onedimerA{black}\right]}\,,
&& Z_3 = \frac{Q\left[\twodimersE{black}{black}\right]}{Q\left[\onedimerA{black}\right]}\,.
\end{align} 

By using a Grassmannian representation of the partition function $Q$ we find
\begin{align}
 &Z_1 = Z_2 = \frac{1}{2}\,, && Z_3 = \frac{1}{\pi}\,.
\end{align} 

\begin{figure}
\begin{center}
\includegraphics[width=6.5in]{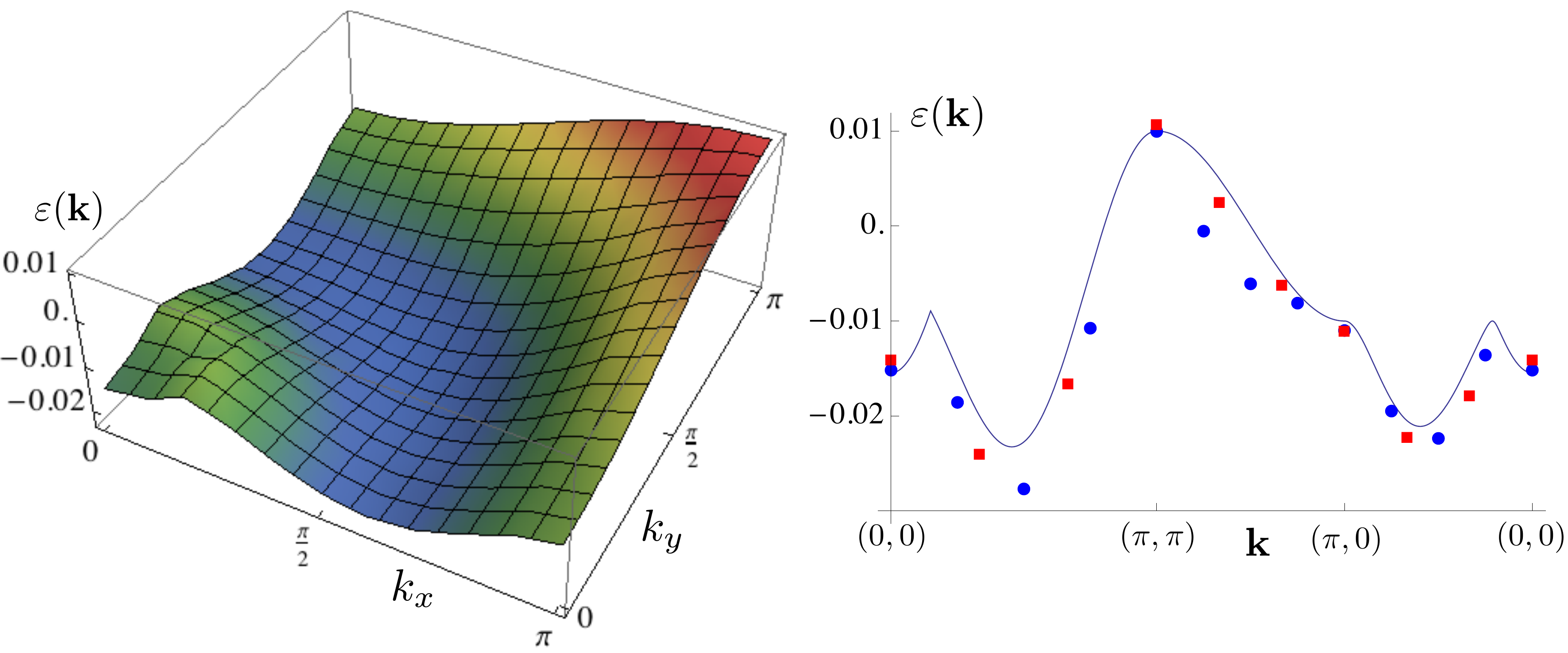}
\end{center}
\caption{The left panel plots the dispersion obtained from diagonalizing the $2\times 2$ matrix in Eq.~(\ref{Hmat}), obtained from perturbation
theory. The right panel compares the perturbative results  (shown as the full line)
with the exact diagonalization of $H_{\rm RK} + H_1$ at $t_1 = t_2 =- t_3 = 0.01$ and $V=J=1$. The latter results are for lattices of size
$6 \times 6$ (red squares) and $8 \times 8$ (blue circles).}
\label{fig:perturb}
\end{figure}

\begin{figure}
\begin{center}
\includegraphics[width=6.5in]{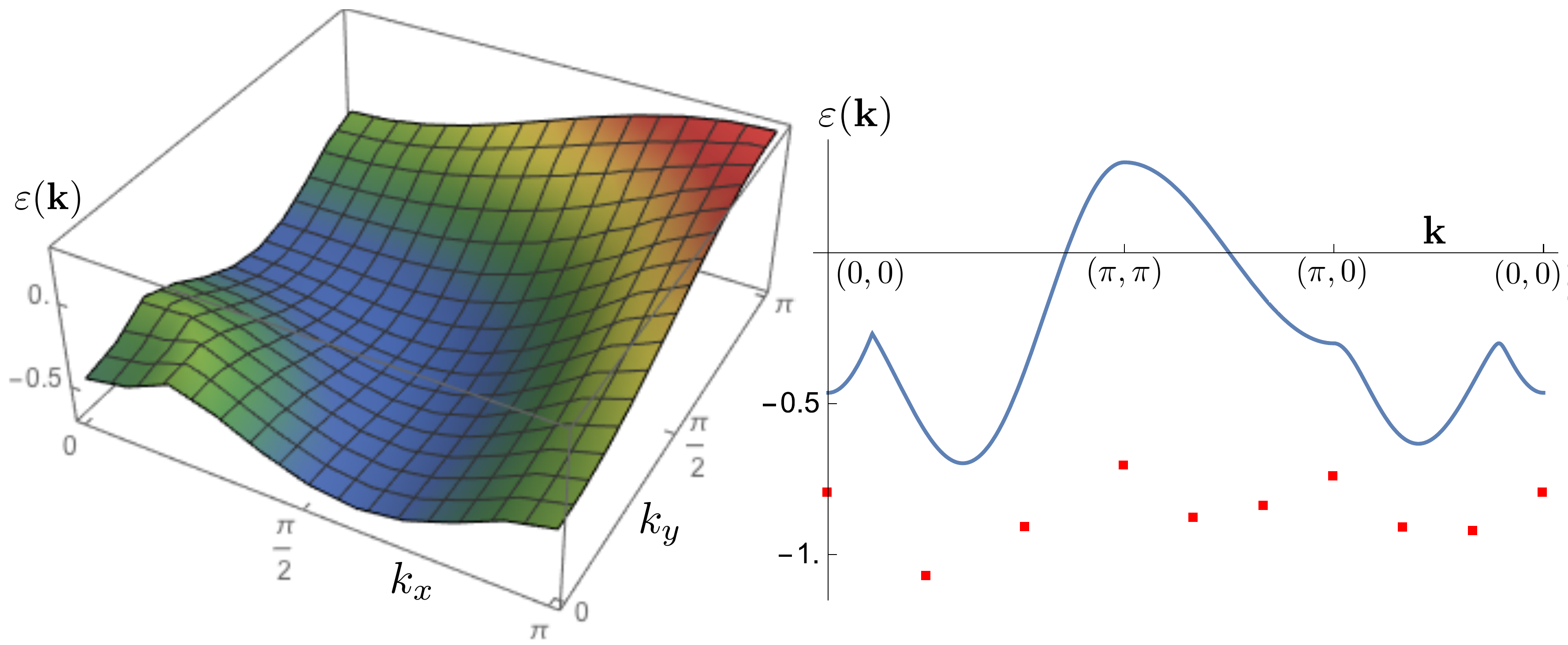}
\end{center}
\caption{As in Fig.~\ref{fig:perturb}, but at $t_1 = t_2 =- t_3 = 0.3$ and $V=J=1$. The exact diagonalization results here are only for the
$6 \times 6$ lattice (red squares).}
\label{fig:perturb2}
\end{figure}

At this point, the problem has been reduced to a ``tight-binding model'' of a single fermion hopping on the links of the square lattice.
So the fermion dispersion and wavefunction is determined by diagonalizing a $2 \times 2$ matrix at each momentum ${\bm k}$:
\beq
H ({\bm k }) = \left( \begin{array}{cc}
-2 t_1 Z_1 \cos(k_y) &  C(k) \\
C^\ast (k) & - 2 t_1 Z_1 \cos(k_x) \end{array} \right) \label{Hmat}
\eeq 
with
\beq C(k) = 
- t_2 Z_2 (1+e^{i k_x})(1 + e^{-i k_y})  - t_3 Z_3 \left[ (1+e^{i k_x})(e^{ik_y} + e^{-2i k_y}) +
(1+e^{-i k_y})(e^{2ik_x} + e^{-ik_x}) \right]
\eeq
We compare this perturbative computation of the quasiparticle energy with the exact diagonalization in Fig.~\ref{fig:perturb}
and find good agreement at $|t_i/J| = 0.01$. At larger values of $|t_i/J|$ the perturbative computation deviates significantly from
the numerics, as shown in Fig.~\ref{fig:perturb2}.

\subsection{Details of the Grassmannian approach}
\begin{figure}
\begin{center}
\includegraphics[width = 0.3 \linewidth]{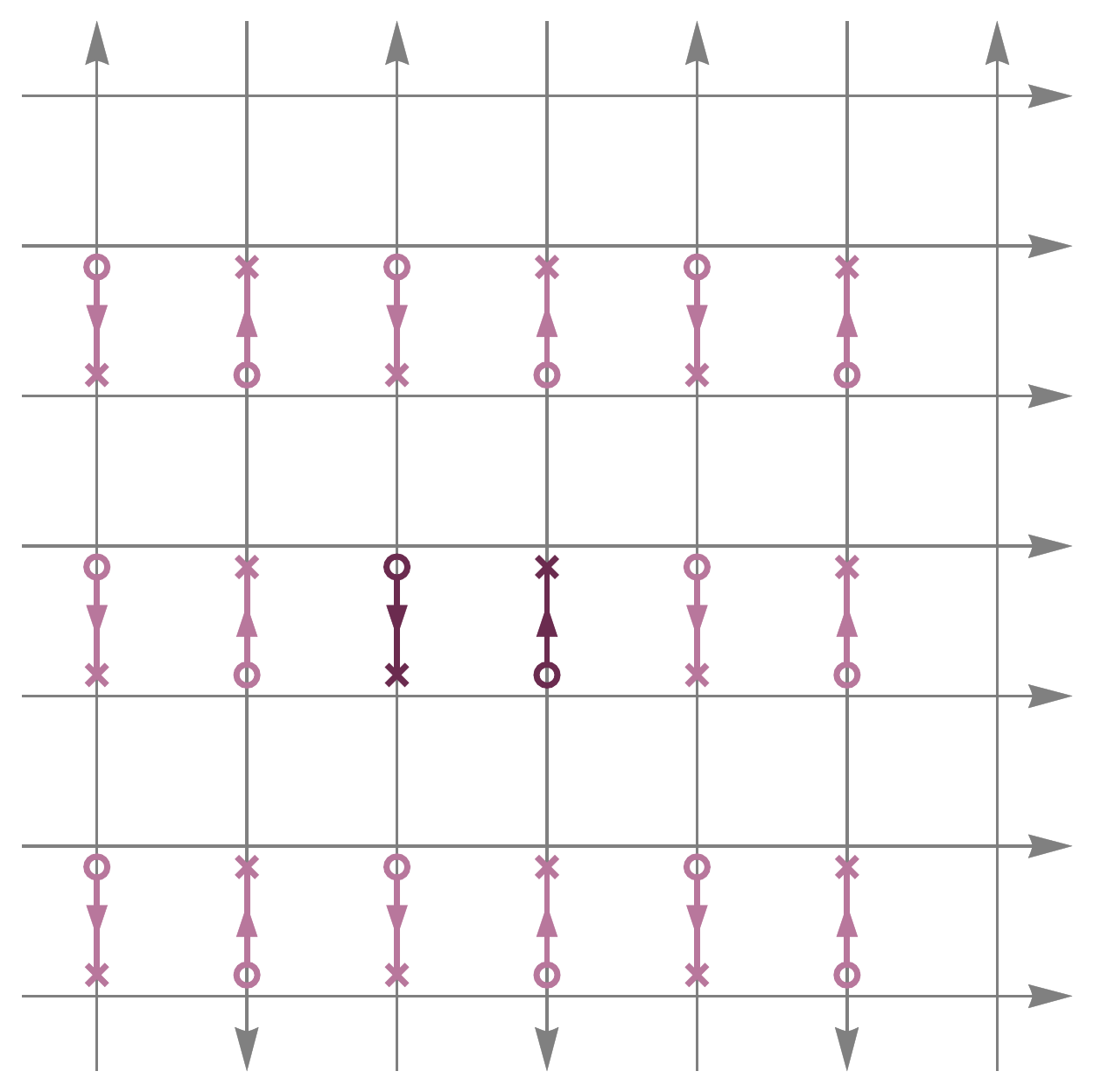}
\includegraphics[width = 0.3 \linewidth]{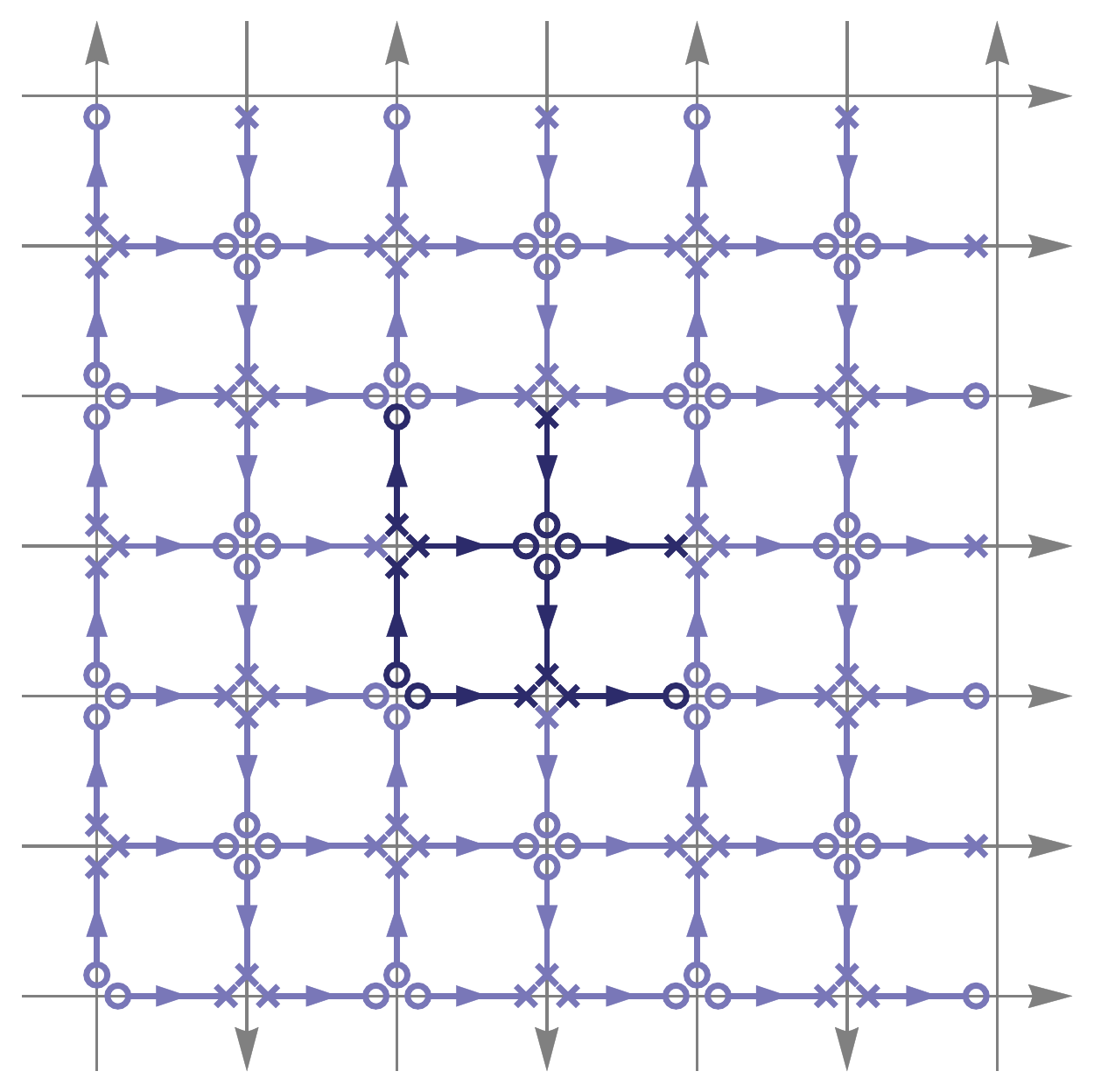}
\end{center}
\caption{\label{fig:pairs} Graphical representation of all terms in the action \eqref{eq:action}. The circle stands for $\eta$ and the cross for $\bar \eta$, and the arrow determines which of the two variables comes first. The bonds emphasized by a darker color form a unit cell that covers the lattice.}
\end{figure}

We use Grassmann variables to compute the function $Q$ \cite{Samuel80}. The partition function with no restriction can be written as
\begin{equation}
Q\left[\zerodimersA\right] = \int \dd \eta \dd \bar \eta\ e^{S(\eta, \bar \eta)}\,,
\end{equation} 
where there is a pair $(\eta \bar \eta)$ of Grassmann variables for every site of the lattice, and the action is quadratic. It contains a pair $(\eta, \bar \eta)$ for every colored bond in fig.~\ref{fig:pairs}, with unit coefficient.  The circle stands for $\eta$ and the cross for $\bar \eta$, and the arrow determines which of the two variables comes first.

When expanding the exponential, only those strings of $\eta$s and $\bar \eta$s that contain exactly one pair for every site will contribute $1$ or $-1$ to the integral. Pairs marked in red in fig.~\ref{fig:pairs} commute with those marked in blue, and every single red pair must appear in the string. On the other hand, only a subset of the blue pairs can appear in the string, and it must form a closed packing dimer covering of the lattice, or else the string will integrate to zero. The non-trivial task is to show that the sign of the contribution will be positive. This is achieved by careful choice of the ordering structure shown by the arrows in  fig.~\ref{fig:pairs}.

Introducing explicit integer coordinates for the lattice we have:
\begin{align}\label{eq:action}
 S = \sum_{m, n} \Big[&
{\color{black!30!ColorData_1_2}\eta_{2m, 2n + 1\phantom{+1}} \bar \eta_{2m, 2n\phantom{+1+1}}} +
{\color{black!30!ColorData_1_2} \eta_{2m + 1, 2n\phantom{+1}} \bar \eta_{2m + 1, 2n + 1}} +\\
\phantom{\sum_{m, n} \Big[}&
{\color{black!30!ColorData_1_1} \eta_{2m, 2n\phantom{+1+1}} \bar \eta_{2m+1,2n\phantom{+1}}} +
{\color{black!30!ColorData_1_1} \bar \eta_{2m, 2n + 1\phantom{+1}} \eta_{2m+1,2n + 1}} +\\
\phantom{\sum_{m, n} \Big[}&
{\color{black!30!ColorData_1_1} \bar\eta_{2m + 1, 2n\phantom{+1}} \eta_{2m+2,2n\phantom{+1}}} +
{\color{black!30!ColorData_1_1} \eta_{2m + 1, 2n + 1} \bar \eta_{2m+2,2n + 1}} +\\
\phantom{\sum_{m, n} \Big[}&
{\color{black!30!ColorData_1_1} \eta_{2m, 2n\phantom{+1+1}} \bar \eta_{2m,2n + 1\phantom{+1}}} +
{\color{black!30!ColorData_1_1} \eta_{2m + 1, 2n + 1} \bar \eta_{2m+1,2n\phantom{+1}}} +\\
\phantom{\sum_{m, n} \Big[}&
{\color{black!30!ColorData_1_1} \bar\eta_{2m, 2n + 1\phantom{+1}} \eta_{2m,2n + 2\phantom{+1}}} +
{\color{black!30!ColorData_1_1} \bar\eta_{2m + 1, 2n + 2} \eta_{2m+1, 2n + 1}}\Big]\,.
\end{align} 

In order to simplify the action, we go to momentum space:
\begin{align}
&\eta_{2m,2n\phantom{+1+1}} = \int \dbar p\, \dbar q\ e^{\ii (pm  + qn)} \psi_{1pq}
&&\bar \eta_{2m,2n\phantom{+1+1}} = \int \dbar p\, \dbar q\ e^{-\ii (pm  + qn)} \bar \psi_{1pq}
\\
&\eta_{2m + 1,2n + 1} = \int \dbar p\, \dbar q\ e^{\ii (pm  + qn)} \psi_{2pq}
&&\bar\eta_{2m + 1,2n + 1} = \int \dbar p\, \dbar q\ e^{-\ii (pm  + qn)} \bar\psi_{2pq}
\\
&\eta_{2m + 1,2n\phantom{+1}} = \int \dbar p\, \dbar q\ e^{\ii (pm  + qn)} \psi_{3pq}
&&\bar\eta_{2m + 1,2n\phantom{+1}} = \int \dbar p\, \dbar q\ e^{-\ii (pm  + qn)} \bar\psi_{3pq}\\
&\eta_{2m,2n + 1\phantom{+1}} = \int \dbar p\, \dbar q\ e^{\ii (pm  + qn)} \psi_{4pq}
&&\bar\eta_{2m,2n + 1\phantom{+1}} = \int \dbar p\, \dbar q\ e^{-\ii (pm  + qn)} \bar\psi_{4pq}
\end{align} 

and we have
\begin{equation}
 S = \int \dbar p\,\dbar q
\begin{pmatrix} \psi_{1pq} & \psi_{2pq} & \psi_{3pq} &\psi_{4pq}\end{pmatrix}
\begin{pmatrix}
0 & 0 & 1 - e^{ \ii p} &   1 - e^{ \ii q}\\
0 & 0 & 1 - e^{-\ii q} &  -(1 - e^{-\ii p})\\
0 & 1 &              0 &                0\\
1 & 0 &              0 &                0\\
\end{pmatrix}
\begin{pmatrix} \bar\psi_{1pq} \\ \bar\psi_{2pq} \\ \bar\psi_{3pq} \\ \bar \psi_{4pq}\end{pmatrix}
\end{equation} 

We may now use the Grassmannian representation to compute ratios of $Q$.

\subsubsection{Normalization}
We know by symmetry that
\begin{equation}
 A_0 \equiv \frac{Q\left[\onedimerA{black}\right]}{Q\left[\zerodimersA\right]} = \frac{1}{4}\,,
\end{equation} 
but it is instructive to obtain this results using the Grassmann approach. Still with reference to fig.~\ref{fig:pairs} and to the discussion in the previous section, we have
\begin{equation}
 A_0 = \expval{\eta_{2m,2n} \bar \eta_{2m, 2n + 1}}\,,
\end{equation} 
where we have chosen the lower left vertical bond, but any other would have worked. We have
\begin{equation}
\begin{split}
 A_0 &= \int \dbar p\, \dbar q\, \dbar r\, \dbar s\ \expval{\psi_{1p q}\bar \psi_{4rs}}\\
 & = \int \dbar p\, \dbar q\ \frac{1 - e^{-\ii q}}{2(2-\cos p -\cos q)} = \frac{1}{4}\,.
\end{split}
\end{equation}
Here we used the fact that for a quadratic action $S = {\psi K \bar \psi}$ we have $ \expval{\psi_i \bar \psi_j} = \left[K^{-1}\right]_{ji}$.

\subsubsection{$Z_1$ and $Z_2$}
We compute
\begin{equation}
 A_1 \equiv \frac{Q\left[\twodimersD{black}{black}\right]}{Q\left[\zerodimersA\right]}.
\end{equation} 

We have
\begin{equation}
\begin{split}
 A_1 & = \expval{\eta_{2m, 2n} \bar \eta_{2m, 2n+1} \eta_{2m + 1, 2n+1} \bar\eta_{2m + 1, 2n}} \\
 & =
   \expval{\eta_{2m, 2n} \bar \eta_{2m, 2n+1}}\expval{\eta_{2m + 1, 2n+1} \bar \eta_{2m + 1, 2n}} + 
   \expval{\eta_{2m, 2n} \bar \eta_{2m + 1, 2n}}\expval{\bar \eta_{2m, 2n+1} \eta_{2m + 1, 2n+1}}\\
 & = \frac{1}{4} \times \frac{1}{4} + \frac{1}{4} \times\frac{1}{4} = \frac{1}{8}\,.
\end{split}
\end{equation} 

Therefore we have
\begin{equation}
 Z_1 = Z_2 = \frac{A_1}{A_0} = \frac{1}{2}
\end{equation} 

\subsubsection{$Z_3$}
We compute
\begin{equation}
 A_3 \equiv \frac{Q\left[\twodimersE{black}{black}\right]}{Q\left[\zerodimersA\right]}.
\end{equation} 

We have
\begin{equation}
\begin{split}
 A_3 & = \expval{\eta_{2m, 2n} \bar \eta_{2m + 1, 2n} \eta_{2m + 2, 2n} \bar \eta_{2m + 2, 2n + 1}} \\
 &=\expval{\eta_{2m, 2n} \bar \eta_{2m + 1, 2n}}\expval{\eta_{2m + 2, 2n} \bar \eta_{2m + 2, 2n + 1}} + \expval{\eta_{2m, 2n} \bar \eta_{2m + 2, 2n + 1} }\expval{\bar \eta_{2m + 1, 2n} \eta_{2m + 2, 2n} } \\
 &= \frac{1}{4}\times\frac{1}{4} + \int \dbar p\, \dbar q\ \frac{e^{-\ii p} (1 - e^{-\ii q}) }{2(2-\cos p -\cos q)} \times \frac{1}{4}\\
&= \frac{1}{4}\times\frac{1}{4} + \left(\frac{1}{\pi} -\frac{1}{4}\right) \times \frac{1}{4}  = \frac{1}{4 \pi}\,,
\end{split}
\end{equation} 
and hence
\begin{equation}
 Z_3 = \frac{A_3}{A_0} = \frac{1}{\pi}\,.
\end{equation}

\end{document}